\documentclass[a4paper,11pt]{article}
\usepackage{jheppub}
\usepackage[utf8]{inputenc}

\usepackage{comment}

\usepackage{dsfont}
\usepackage{simplewick}

\usepackage{mathtools,amsmath,amssymb}
\usepackage{graphicx}

\usepackage{lmodern}
\usepackage[T1]{fontenc}
\usepackage{microtype}

\usepackage{hyperref}

\usepackage{xspace}

\usepackage{array} 

\usepackage{subfigure}

\usepackage{pgf}
\usepackage{tikz}
\usetikzlibrary{shapes,arrows,automata,positioning}
\usepackage{tikz-3dplot}

\newcommand{\YM}{{\mathrm{\scriptscriptstyle YM}}}

\usepackage{xspace}
\usepackage{tikz}
\usetikzlibrary{shapes}
\usetikzlibrary{decorations.markings}

\tikzset{->-/.style={decoration={
  markings,
  mark=at position .5 with {\arrow{>}}},postaction={decorate}}}
  
\usepackage{placeins}

\DeclareMathOperator{\phaneq}{\phantom{{}=}}

\newcommand{\half}{{\tfrac{1}{2}}}

\newcommand{\e}{\operatorname{e}}
\newcommand{\de}{\operatorname{d}\!}

\newcommand{\abs}[1]{{ | {#1} | }}

\newcommand{\calB}{\mathcal{B}}

\newcommand{\cO}{\mathcal{O}}

\newcommand{\eqndot}{\, . }
\newcommand{\eqncom}{\, , }

\newcommand{\Zbar}{\bar{Z}}
\newcommand{\Zc}{Z^{\text{cl}}}
\newcommand{\Zbarc}{\Zbar^{\text{cl}}}

\DeclareMathOperator{\tr}{tr}

\DeclareMathOperator{\idm}{\mathds{1}}

\renewcommand{\dbinom}[2]{\Big(\!\genfrac{}{}{0pt}{1}{#1}{#2}\!\Big)}

\makeatletter
\renewcommand*\env@matrix[1][\arraystretch]{%
  \edef\arraystretch{#1}%
  \hskip -\arraycolsep
  \let\@ifnextchar\new@ifnextchar
  \array{*\c@MaxMatrixCols c}}
\makeatother

\makeatletter
\def\mmatrix{\@ifnextchar[{\@amwith}{\@mwithout}}
\def\@amwith[#1]{\@ifnextchar[{\@mwwith[#1]}{\@mwith[#1]}}
\def\@mwwith[#1][#2]#3{
\begingroup\setlength{\arraycolsep}{#2pt}
    \begin{pmatrix}[#1]
        #3
    \end{pmatrix}
\endgroup
}
\def\@mwith[#1]#2{
    \begin{pmatrix}[#1]
        #2
    \end{pmatrix}
}
\def\@mwithout#1{
    \begin{pmatrix}
        #1
    \end{pmatrix}
}
\makeatother

\usepackage[textsize=footnotesize]{todonotes}

\makeatletter
\DeclareRobustCommand*{\bfseries}{%
  \not@math@alphabet\bfseries\mathbf
  \fontseries\bfdefault\selectfont
  \boldmath
}

\makeatother

\usepackage{lipsum}
\makeatletter
\if@titlepage
  \renewenvironment{abstract}{%
      \titlepage
      \null\vfil
      \@beginparpenalty\@lowpenalty
      \begin{center}%
        \bfseries \abstractname
        \@endparpenalty\@M
      \end{center}}%
     {\par\vfil\null\endtitlepage}
\else
  \renewenvironment{abstract}{%
      \if@twocolumn
        \section*{\abstractname}%
      \else
        \small
        \begin{center}%
          {\bfseries \abstractname\vspace{-.5em}\vspace{\z@}}%
        \end{center}%
        \quotation
      \fi}
      {\if@twocolumn\else\endquotation\fi}
\fi
\makeatother

\renewcommand{\digamma}{\Psi}
        
\DeclareMathOperator{\cder}{D}
\newcommand{\scal}{\phi}
\newcommand{\scalc}{\scal^{\text{cl}}}
\newcommand{\scalq}{\tilde{\scal}}
\newcommand{\ferm}{\psi}
\newcommand{\aferm}{\bar{\ferm}}

\newcommand{\comm}[2]{[#1,#2]}

\newcommand{\soe}{\text{SO}(3)_{E}} 
\newcommand{\soc}{\text{SO}(3)_{C}}
\newcommand{\fscal}{q}
\newcommand{\afscal}{\bar q}
\newcommand{\fferm}{\chi}
\newcommand{\afferm}{\bar \chi}

\newcommand{\spec}{\text{c.}}
        
\title{Two-point functions in AdS/dCFT and the boundary conformal bootstrap equations}
\author{Marius de Leeuw, Asger C. Ipsen, Charlotte Kristjansen, Kasper E.\ Vardinghus and Matthias Wilhelm}

\begin{document}

\begingroup\parindent0pt
\begin{flushright}\footnotesize
\end{flushright}
\vspace*{4em}
\centering
\begingroup\LARGE
\bf
Two-point functions in AdS/dCFT and the boundary conformal bootstrap equations 
\par\endgroup
\vspace{2.5em}
\begingroup\large{\bf Marius de Leeuw, Asger C.\ Ipsen, Charlotte Kristjansen,\\ Kasper E.\ Vardinghus and Matthias Wilhelm}
\par\endgroup
\vspace{1em}
\begingroup\itshape
Niels Bohr Institute, Copenhagen University,\\
Blegdamsvej 17, 2100 Copenhagen \O{}, Denmark\\

\par\endgroup
\vspace{1em}
\begingroup\ttfamily
deleeuwm@nbi.ku.dk, asgercro@nbi.ku.dk, kristjan@nbi.ku.dk, kasper.vardinghus@nbi.ku.dk, matthias.wilhelm@nbi.ku.dk \\
\par\endgroup
\vspace{2.5em}
\endgroup

\begin{abstract}
\noindent
We calculate the leading contributions to the connected  two-point functions of protected scalar operators in the defect version of $\mathcal{N}=4$ SYM theory which is dual to the D5-D3 probe-brane system with $k$ units of background gauge field flux. This involves several types of two-point functions which are vanishing in the theory without the defect, such as two-point functions  of operators of unequal conformal dimension. We furthermore exploit the operator product expansion (OPE) and the boundary operator expansion (BOE), which form the basis of the boundary conformal bootstrap equations, to extract conformal data both about the defect CFT and about ${\cal N}=4$ SYM theory without the defect.
From the knowledge of the one- and two-point functions of the defect theory, we extract certain structure constants of ${\cal N}=4$ SYM theory using
the (bulk) OPE and constrain certain bulk-bulk-to-boundary
couplings using the BOE. The extraction of the former relies on a non-trivial, polynomial $k$ dependence of the one-point functions, which we explicitly demonstrate. In addition, it requires the knowledge of the one-point functions of SU$(2)$ descendant operators, which we likewise explicitly determine.
  
\end{abstract}


\thispagestyle{empty}

\newpage
\hrule
\tableofcontents
\afterTocSpace
\hrule
\afterTocRuleSpace

\section{Introduction}

Introducing boundaries or defects in conformal field theories (CFTs) leads to interesting new structures~\cite{Cardy:1984bb} and constitutes a simple path towards studying various types of symmetry breaking. What is more, the introduction of defects or boundaries makes the  theories more adapt for studying realistic physical systems.

Several types of correlation functions which vanish in the absence of a defect become non-trivial when a defect is introduced.
For instance, one-point functions of bulk operators can be non-zero and so can two-point functions involving operators of unequal conformal dimension. Moreover, two-point functions are not completely fixed by symmetries.
Accordingly, defect conformal field theories require a larger amount of conformal data for their
specification. Among this data are the one-point functions of bulk operators and the two-point functions between bulk and boundary operators.
The bootstrap program 
for conformal field theories can 
be extended to defect conformal field theories (dCFTs) as well, where additional bulk-to-boundary crossing relations for two-point functions come into play~\cite{Liendo:2012hy,Gliozzi:2015qsa,Billo:2016cpy}.

An interesting 4D dCFT can be constructed starting from ${\cal N}=4$ SYM theory with gauge group U$(N)$ by introducing a codimension-one defect,
say at $x_3=0$~\cite{DeWolfe:2001pq}. The defect supports a fundamental hypermultiplet of fields which self-interact and
interact with the bulk fields of ${\cal N}=4$ SYM theory in such away that half of the supersymmetries of ${\cal N}=4$ SYM theory are
preserved and the complete symmetry group of the theory is OSp$(4|4)$. A particular version of the theory, where the features
of a dCFT  are visible already at tree level, can be obtained by assigning three of the six scalar fields of ${\cal N}=4$
SYM theory a non-vanishing and space-time-dependent vacuum expectation value (vev) on one side of the defect, $x_3>0$.
This dCFT has a holographic dual consisting of a D5-D3 probe-brane system where the D5 brane has geometry
$AdS_4\times S^2$ and where a certain background gauge field has $k$ units of magnetic flux on the $S^2$~\cite{Karch:2000gx,Constable:1999ac}.
The latter statement is equivalent to the statement that $k$ of the $N$ D3 branes of the usual $AdS_5\times S^5$ set-up terminate on the D5 brane.
The vevs of the
scalars in the field theory reflect the so-called fuzzy funnel solution of the probe-brane system and 
result in the gauge group of ${\cal N}=4$ SYM theory being (broken) U$(N)$ on one side of the defect, $x_3>0$, and
U$(N-k)$ on the other side, $x_3<0$.

In our previous work, we have set up the program for carrying out perturbative calculations in the dCFT above, which
required diagonalising a highly involved mass matrix as well as devicing a way to work with space-time-dependent  mass parameters~\cite{Buhl-Mortensen:2016jqo,Buhl-Mortensen:2016pxs}. This has opened a vast arena for the calculation of all 
possible types of correlation functions of an interesting 4D  dCFT. We have already devoted some attention to one-point functions
of the theory, where firstly we found interesting connections to 
integrability~\cite{deLeeuw:2015hxa,Buhl-Mortensen:2015gfd,deLeeuw:2016umh,Buhl-Mortensen:2017ind}
and secondly were able to perform a non-trivial quantum check of the gauge-gravity correspondence in a situation 
where both supersymmetry and conformal symmetry were partly broken~\cite{Buhl-Mortensen:2016jqo,Buhl-Mortensen:2016pxs,Buhl-Mortensen:2017ind}.  

In the present paper, our focus will be on two-point functions of bulk operators.  The general form of such correlators can be constrained by symmetry arguments~\cite{Cardy:1984bb}. We demonstrate how the predicted behaviour emerges from our previously derived Feynman rules and give the explicit expressions for the correlators.
Moreover, we exploit the conformal boundary bootstrap equations~\cite{Liendo:2012hy,Gliozzi:2015qsa,Billo:2016cpy,Liendo:2016ymz}, or, more  precisely, the (bulk) operator product expansion (OPE) and boundary operator expansion (BOE) to extract additional conformal data  both about the dCFT and about ${\cal N}=4$ SYM theory without the defect.%
\footnote{Note that throughout this paper we are also referring to the defect at $x_3=0$ as boundary and to the regions of space-time with $x_3\neq0$ as bulk; 
in particular, these notions of boundary and bulk should not be confused with those occurring in the context of AdS$_5$ in the gauge-gravity correspondence. In addition, strictly speaking, we are looking at an interface rather than a boundary; however, the former can always be mapped to the latter; see for instance \cite{Gaiotto:2008sa}.}
Using as input the two-point functions of the dCFT in combination with our previously derived one-point functions, we obtain structure constants of the theory without the defect from the bulk OPE, and we constrain bulk-to-boundary 
couplings of the dCFT  from the BOE. 
The exploitation of the former OPE requires the knowledge of the one-point functions of certain descendant operators\footnote{The operators in question are descendants in the SU(2) sense and not conformal descendants.} 
which were not known  before but which we compute here in full generality.  
It also requires a rewriting of the one-point functions as polynomials in $k$, which we likewise provide. In order to not only constrain but explicitly determine the
bulk-to-boundary couplings, a detailed derivation of the operator content and the interactions of the boundary theory is needed. We explicitly derive the spectrum of
boundary operators but given its complexity we postpone the analysis of the interactions to future work.
 
We start in section~\ref{sec:dCFT} by reviewing the defect version of ${\cal N}=4$ SYM theory and the constraints from conformal symmetry on a dCFT, which will be the
basis of our analyses. We move on to calculating, in section~\ref{sec:Two-point}, a series  of bulk two-point functions of BPS operators. 
Section~\ref{sec:OPE} is concerned with the above mentioned data mining using the bulk OPE as well as the BOE and section~\ref{sec:Conclusion} contains our conclusion. 
A number of derivations are relegated to appendices. 
Hence, in appendix~\ref{App:Descendants} we derive a closed expression for the one-point functions of descendant operators, in appendix~\ref{App:Rewriting} we present the rewriting of the one-point functions as polynomials in $k$ and  in appendix~\ref{App:Fuzzy} we have collected a number of useful identities for fuzzy spherical harmonics that arise in the diagonalisation of the mass matrix and hence play an important role in the evaluation of correlations functions. 
Finally, in appendix~\ref{app: boundary operators}  we present the spectrum of gauge-invariant boundary operators of the theory.

\section{The Defect Theory\label{sec:dCFT}}

\subsection{Action and propagators}
\label{sec:action}


We consider four-dimensional $\mathcal{N}=4$ SYM theory in the bulk interacting with a defect of codimension one situated at $x_3=0$ \cite{DeWolfe:2001pq, Erdmenger:2002ex}. The action of $\mathcal{N}=4$ SYM theory in our conventions reads
\begin{multline}
 \label{eq: SYM-action}
  S_{{\cal N}=4}=\frac{2}{g_\YM^2}\int \de^4x\tr\biggl[ -\frac{1}{4}F_{\mu\nu}F^{\mu\nu}-\frac{1}{2}\cder_\mu\scal_i\cder^\mu\scal_i\\+\frac{i}{2}\aferm\Gamma^\mu\cder_\mu\ferm +\frac{1}{2}\aferm\Gamma^i\comm{\scal_i}{\ferm}+\frac{1}{4}\comm{\scal_i}{\scal_j}\comm{\scal_i}{\scal_j}\biggr]\eqncom
\end{multline}
where $F_{\mu\nu}$ is the field strength, $\cder_\mu$ denotes the covariant derivatives and $\Gamma$ are the ten-dimensional gamma matrices describing the couplings of the fermions $\ferm$ to the gauge field and the six real scalars $\phi_i$, $i=1,\ldots,6$.

A solution to the classical equations of motion of the system is given by assigning a non-vanishing and $x_3$-dependent vev to three of the six real scalars, say $\phi_1$, $\phi_2$ and $\phi_3$, while having all other classical fields vanish \cite{Constable:1999ac}.
The resulting equations of motions can be solved by
\begin{equation}
 \label{eq: classical solution}
 \langle\phi_i\rangle_{\text{tree}}= \scalc_i=-\frac{1}{x_3} t_i\oplus 0_{(N-k)\times(N-k)}\eqncom \qquad x_3>0 \eqncom
\end{equation}
where $i=1,2,3$ and $t_i$ are the generators of the $k$-dimensional irreducible representation of the SU(2) Lie algebra.
As a consequence, the gauge group U$(N)$ is broken for $x_3>0$ and can be taken to be U$(N-k)$ for $x_3<0$.
The expression~(\ref{eq: classical solution}) also solves the Nahm equations \cite{Nahm:1979yw}.

The action of the complete system includes a three-dimensional action describing the defect fields and their interaction with the bulk fields. The three-dimensional 
action was worked out for the $k=0$ case in~\cite{DeWolfe:2001pq}.  
For the present calculation of connected two-point functions of bulk operators at leading order, the defect action does however not play a role.%
\footnote{The defect action is expected to play a role for higher loop corrections. For a discussion of this point, we refer to
\cite{Buhl-Mortensen:2016jqo}. It obviously will also play a role for correlators involving defect fields.}

In order to do perturbative calculations in the resulting dCFT, we expand the fields around the classical solution:
\begin{equation}
 \phi_i=\scalc_i+\scalq_i\eqncom \qquad i=1,2,3\eqndot
\end{equation}
After gauge fixing, this leads to $x_3$-dependent mass terms for the gauge fields, scalars, fermions and ghosts, which are non-diagonal in colour space as well as in flavour space.
Moreover, new cubic interaction terms arise. The expanded action is explicitly given in \cite{Buhl-Mortensen:2016jqo}.
There, we have also diagonalised the mass matrix using spherical harmonics of the fuzzy sphere.
The resulting eigenvalues follow an intricate pattern when partially expressed in terms of 
\begin{equation}
 \label{eq: def nu}
  \nu=\sqrt{m^2+\frac{1}{4}}\eqncom
\end{equation}
and are explicitly given in table \ref{tab:spectrum}.

\begin{table}[t]
\centering
\begin{tabular}{c|c|c|c}
Multiplicity & $\nu(\scalq_{4,5,6},A_{0,1,2},c)$ & $m(\psi_{1,2,3,4})$ & $\nu(\scalq_{1,2,3},A_3)$ \\ \hline
$\ell +1$ & $\ell+\frac{1}{2}$ & $\ell$ & $\ell-\frac{1}{2}$ \\
$\ell$ & $\ell+\frac{1}{2}$ & $\ell+1$ & $\ell+\frac{3}{2}$ \\
$(k+1)(N-k)$ & $\frac{k}{2}$ & $\frac{k-1}{2}$ & $\frac{k-2}{2}$ \\
$(k-1)(N-k)$ & $\frac{k}{2}$ & $\frac{k+1}{2}$ & $\frac{k+2}{2}$ 
\end{tabular}
\caption{\label{tab:spectrum} Masses and $\nu$'s of the modes propagating only for $x_3>0$, with $\ell=0,\ldots,k-1$ \cite{Buhl-Mortensen:2016pxs,Buhl-Mortensen:2016jqo}.
In addition, $(N-k)^2$ massless modes exist that propagate on both sides of the defect.
} 
\end{table}

Due to the $x_3$ dependence of the classical solution \eqref{eq: classical solution}, all masses  are accompanied by a factor of $1/x_3$.
As a result, the propagators of the massive modes take the form of propagators in an effective $AdS_4$ space \cite{Nagasaki:2011ue,Buhl-Mortensen:2016pxs,Buhl-Mortensen:2016jqo}. 
For instance, we have for scalars with mass parameter $\nu$\footnote{For $\nu = -1/2$, the
meaning of the right-hand side is defined by the limit $\nu \to -1/2$. In particular,
one should set $_2F_1(-1,0;0;-\xi^{-1}) = 1 + \frac{1}{2}\xi^{-1}$.}
\begin{equation}
\begin{aligned}
  K^\nu(x,y) &= \frac{g_\YM^2}{2}\frac{K_{\text{AdS}}(x,y)}{x_3y_3}=\frac{g_\YM^2}{16\pi^2}\frac{1}{\dbinom{2\nu+1}{\nu+\half}}
\frac{_2F_1(\nu-\half,\nu+\half;2\nu+1;-\xi^{-1})}{(1+\xi)\xi^{\nu+\half}}\frac{1}{x_3y_3} \eqncom
\end{aligned}  
\label{eq:Knu}
\end{equation}
where $\xi$ is the conformal ratio
\begin{equation}\label{def:xi}
\xi=\frac{|x -y|^2}{4 x_3 y_3}\eqndot
\end{equation}
The AdS propagator can for instance be found in \cite{Allen:1985wd,Camporesi91}.%
\footnote{In our previous works \cite{Buhl-Mortensen:2016pxs,deLeeuw:2016vgp,Buhl-Mortensen:2016jqo}, we were using an integral representation for the AdS propagator \cite{Liu:1998ty} that facilitates its regularisation. In the present calculation, regularisation is not necessary as all quantities are manifestly finite.}

A subtle point is related to the boundary conditions of the massless modes in U$(N)$ that are not present in U$(N-k)$.
First, there exists a massless bosonic mode for $\ell=1$, shown in the last column of the first row of table \ref{tab:spectrum}. Since this mode is related to the massive fermionic and bosonic modes in the first row via supersymmetry, this mode is restricted to $x_3>0$ with Dirichlet boundary conditions at the defect.
Second, there is a massless supermultiplet with $\ell=0$ , shown in the first row in table \ref{tab:spectrum}.
The brane construction of the dCFT suggests that the gauge group is U$(N-k)$ for $x_3<0$.
The $\ell = 0$ modes  are thus restricted to $x_3 > 0$, with appropriate boundary conditions at the defect.
It turns out that there are two possibilities compatible with supersymmetry, cf.\ \cite{Gaiotto:2008sa}: a) we can choose Dirichlet boundary conditions for the first column and Neumann boundary conditions for the last column, resulting in a gauge group U$(N-k)$ at $x_3=0$, or b) we can choose Neumann boundary conditions for the first column and Dirichlet boundary conditions for the last column, resulting in a gauge group U$(N-k)\times$U$(1)$ at $x_3=0$. It is gratifying to see that the natural extension of our
 expressions from $\ell > 0$ to $\ell = 0$ automatically gives the first kind of supersymmetric boundary conditions.\footnote{Indeed
$K^{\nu=1/2}$ ($K^{\nu=-1/2}$) satisfies Dirichlet (Neumann) boundary conditions.}
To get the second kind, on would have to reverse the boundary conditions by hand,
and this would also be  consistent in so far as the results of this paper are concerned (but the explicit one- and two-point
functions would be different). 
It would be interesting to see whether this continues to hold when considering a larger class of observables or higher loop orders.

\subsection{Operator product and boundary operator expansion}\label{sec:OPEandBOE}

Let us briefly review the consequences of conformal symmetry in the presence of a defect or boundary, cf.\ e.g.\ \cite{McAvity:1995zd, Gliozzi:2015qsa, Billo:2016cpy, Liendo:2012hy}.

In a usual CFT, one-point functions of composite operators $\cO_i$ are vanishing and the space-time dependence of two-point functions is completely fixed by the scaling dimensions of the operators $\Delta_i$, which have to be equal. In particular, one can define an `orthonormal' set of operators by diagonalising the matrix $M_{ij} = \langle\cO_i\cO_j\rangle|_{|x-y|=1}$ of two-point functions.
Conformal symmetry further fixes the three-point function up to one constant, the structure constant $\lambda_{i j k}$, which appears in the operator product expansion (OPE)
\begin{align}\label{eq:GeneralOPE}
\mathcal{O}_i(x) \mathcal{O}_j(y) = \frac{M_{ij}}{|x-y|^{\Delta_{i}+\Delta_{j}}} + \sum_k \frac{\lambda_{i j}{}^{k}} {|x-y|^{\Delta_{i}+\Delta_{j}-\Delta_k} }\,C(x-y,\partial_y)\mathcal{O}_k(y)\eqncom
\end{align}
where the sum over $k$ runs over conformal primaries and the differential operator $C$ in \eqref{eq:GeneralOPE} accounts for the presence of descendants. It is easy to see that the indices on $\lambda$ are raised and lowered with the two-point function matrix $M$. The normalization of $C$ is such that 
$C(x-y,\partial_y)=1+{\cal O}(x-y)$.
Starting from four-point functions, a non-trivial (space-time) dependence on conformal cross-ratios can occur. 
However, the scaling dimensions $\Delta_i$ and the structure constants $\lambda_{i j k }$, the so-called conformal data, completely determine all four- and higher-point functions  via consecutive applications of the OPE.
Equating the different ways to apply the OPE leads to powerful consistency conditions, the so-called bootstrap equations.

A CFT with a boundary or defect has a richer structure.
Here, also one-point functions of composite operators $\cO_i$ can be non-vanishing. Conformal symmetry and the scaling dimension $\Delta_i$ of the operator fix the one-point functions up to a 
constant $a_i$ \cite{Cardy:1984bb}:%
\footnote{We adopt here the convention usually used in the boundary bootstrap program, which involves an explicit factor of 2 in the space-time dependence.}
\begin{equation}
 \langle \mathcal{O}_i(x)\rangle=\frac{a_i}{(2x_3)^{\Delta_{i}}}\eqndot
\end{equation}
Thus, one-point functions in a dCFT exhibit a complexity similar to three-point functions in a CFT.
Two-point functions in a dCFT can be non-vanishing also for operators of unequal scaling dimensions and are fixed to be of the form
\begin{align}
\langle \mathcal{O}_i(x) \mathcal{O}_j(y) \rangle = \frac{f(\xi)}{(2x_3)^{\Delta_i}(2y_3)^{\Delta_j}}\eqncom
\end{align}
where $f(\xi)$ is a function of the conformal ratio \eqref{def:xi}.

The two-point function tends to  the one in the absence of the defect if the distance to the defect is large compared to the distance between the points:
\begin{equation}
\lim_{z_3\to\infty}\langle \mathcal{O}_i(x+z) \mathcal{O}_j(y+z) \rangle=\frac{M_{ij}}{|x-y|^{\Delta_{i}+\Delta_{j}}}
\eqndot
\end{equation}
Then, since the OPE expansion is around the point $\xi=0$, the OPE of the operators in the bulk \eqref{eq:GeneralOPE} is unchanged.
Using this OPE, we can express the two-point function in terms of $\Delta$, $\lambda$ and $a$ as \cite{Liendo:2012hy}
\begin{align}
\label{eq: OPE for two-point}
f(\xi) = \xi^{-\frac{\Delta_i+\Delta_j}{2}}\left[M_{ij}+ \sum_k \lambda_{i j}{}^{k} a_{k} F_{\text{bulk}}(\Delta_k,\Delta_i-\Delta_j,\xi)\right]\eqncom
\end{align}
where the bulk conformal block is given by
\begin{align}
 F_{\text{bulk}}(\Delta,\delta\Delta,\xi) = \xi^{\frac{\Delta}{2}}{}_2F_1(\tfrac{1}{2}(\Delta+\delta\Delta),\tfrac{1}{2}(\Delta-\delta\Delta);\Delta-1;-\xi)\eqndot
\end{align}
Thus, two-point functions in the dCFT exhibit a complexity similar to the one of four-point functions in a usual CFT.

A further feature of dCFTs is the existence of boundary or defect operators.
Since the theory on the defect is a usual CFT in one dimension less, the space-time dependence of the boundary-boundary two-point functions is completely determined by the scaling dimensions. Moreover, the boundary operators posses an OPE fixed by structure constants $\hat{\lambda}$, which allows to construct all higher-point functions of boundary operators.%
\footnote{Strictly speaking, one can even eliminate the bulk structure constants $\lambda$ from the list of conformal data by expressing them in terms of the boundary structure constants $\hat{\lambda}$ and the bulk-to-boundary couplings $\mu$ defined below, see for instance \cite{Gliozzi:2015qsa,Billo:2016cpy}. This is, however, not the approach we will be taking here.}

\begin{figure}[t]
 \centering
 
$\begin{aligned}
\includegraphics[width=0.25\textwidth,keepaspectratio]{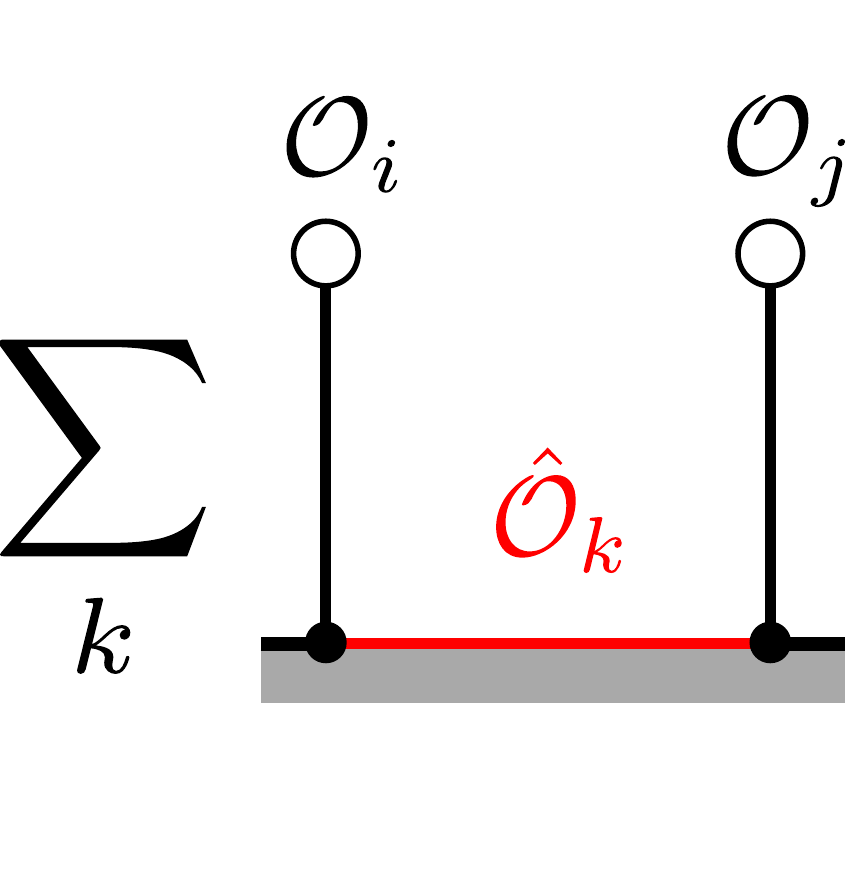}
\end{aligned} =\langle\mathcal{O}_i\mathcal{O}_j\rangle=
\begin{aligned}
\includegraphics[width=0.25\textwidth,keepaspectratio]{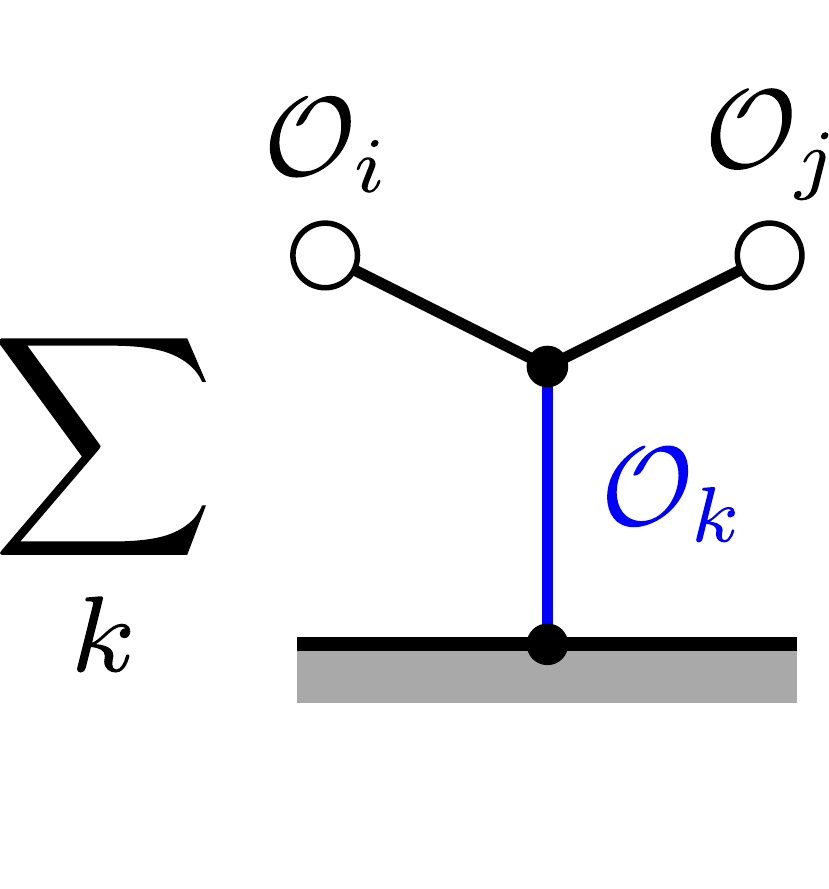}
\end{aligned} $

\caption{The two-point function of two operators $\mathcal{O}_i$ and $\mathcal{O}_j$ can be expressed in two ways: via the BOE (left) and the OPE (right).} 
\label{fig: bootstrap equation}
\end{figure}

However, we can also have non-vanishing two-point functions between a bulk and a boundary operator.
Via conformal symmetry, these are fixed to be of the form 
\begin{equation}
 \langle \mathcal{O}_i(x) \hat{\mathcal{O}}_j(\vec{y}) \rangle =\frac{\mu_{ij}}{(2x_3)^{\Delta_{i}-\Delta_{j}}|x-(\vec{y},0)|^{2\Delta_{j}}}\eqncom
\end{equation}
where we use the notation $\vec{y}=(y_0,y_1,y_2)$ for the coordinates on the defect.
The coefficients $\mu_{ij}$ originate from the expansion of the bulk operators in terms of boundary operators, the boundary operator expansion (BOE): 
\begin{equation}
 \cO_i(x)=\sum_j\frac{\mu_{i}{}^{j}}{(2x_3)^{\Delta_i-\Delta_j}}\hat{C}(x_3,\vec{\partial})\hat\cO_j(\vec{x})\eqncom
\end{equation}
where $\mu_{i\hat{\idm}}\equiv a_{i}$ and where the differential operator $\hat{C}$ accounts for the decendants on the boundary and is normalized such that
$\hat{C}=1+{\cal O}(x_3^2)$.
 The second index on $\mu$ is raised and lowered by $\hat{M}$, the matrix of two-point functions of boundary operators.
The BOE provides us with a second way to express the bulk-bulk two-point function:
\begin{align}
\label{eq: BOE for two-point}
f(\xi) = 
a_{i}a_{j} + \sum_k \mu_{i}{}^{k}\mu_{jk} F_{\text{bdy}}(\Delta_k,\xi)\eqncom
\end{align}
where the boundary conformal block is given by
\begin{align}
\label{eq: boundary conformal block}
 F_{\text{bdy}}(\Delta,\xi) = \xi^{-\Delta}{}_2F_1(\Delta,\Delta-1;2\Delta-2;-\xi^{-1})\eqndot
\end{align}
Note that the second term in \eqref{eq: BOE for two-point} stems from the connected two-point function while the first term stems from the disconnected product of the one-point functions.

Equating \eqref{eq: OPE for two-point} and \eqref{eq: BOE for two-point} as illustrated in figure \ref{fig: bootstrap equation} leads to bootstrap equations in the presence of a defect that can be used to constrain the conformal data \cite{Liendo:2012hy,Gliozzi:2015qsa,Billo:2016cpy,Liendo:2016ymz,Hogervorst:2017kbj}.
In the following, however, we take a different route: 
we will be explicitly calculating bulk-bulk two-point functions and use \eqref{eq: OPE for two-point} and \eqref{eq: BOE for two-point} to extract conformal data.

\section{Two-point functions\label{sec:Two-point}}

\begin{figure}[t]
\centering
  \includegraphics{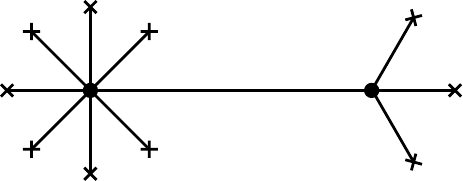}%
  \label{tree}
\caption{The leading contribution to the connected two-point function of scalar operators.   The operators are represented by a dot and a cross symbolises the insertion of the classical solution.\label{fig: two-point}}
\end{figure}

In this section, we calculate the leading contribution to the connected two-point functions of BPS operators built from complex scalar fields. This amounts to evaluating a Feynman diagram of the type depicted in figure~\ref{fig: two-point}.

\paragraph{Scalar fields}  We define complex combinations of the scalar fields
as follows:
\begin{align}
X=\phi_1+i\phi_4,\hspace{0.5cm} Z=\phi_3+i\phi_6.
\end{align}
There are three different cases of Wick contractions between the scalar fields. 
In what follows, we will only need the contraction rules for the fields in the $k\times k$ block. Following our work~\cite{Buhl-Mortensen:2016jqo},
these fields need to be expanded in terms of fuzzy spherical harmonics $\hat{Y}^m_\ell$ as $Z=(Z)_{\ell m}\hat{Y}^m_\ell$, $\bar{Z}=(\bar{Z})_{\ell m}\hat{Y}^m_\ell$, etc. The Wick contractions can then be worked out following \cite{Buhl-Mortensen:2016pxs,Buhl-Mortensen:2016jqo} to give
\begin{align}\label{eq: Contractions}
\langle Z_{\ell m}(x) Z_{\ell' m'}(y)\rangle &= \delta_{\ell \ell'}\delta_{m + m',0} \frac{g_\YM^2}{16\pi^2}\frac{(-1)^{m'} }{x_3 y_3}
\frac{\, _2F_1(\ell,\ell+1;2 \ell+2;-\xi^{-1} )}{\binom{2\ell+1}{\ell+1}\xi^{\ell+1}}
\frac{\xi}{\xi +1}\eqncom\nonumber\\
\langle Z_{\ell m}(x) \bar{Z}_{\ell' m'}(y)\rangle &= \delta_{\ell \ell'}\delta_{m + m',0}\frac{g_\YM^2}{16\pi^2} \frac{(-1)^{m'} }{x_3 y_3}
\frac{\, _2F_1(\ell,\ell+1;2 \ell+2;-\xi^{-1} )}{\binom{2\ell+1}{\ell+1}\xi^{\ell+1}}\eqncom\\ 
\langle Z_{\ell m}(x) X_{\ell' m'}(y)\rangle &= \delta_{\ell \ell'} \frac{[t^{(2\ell+1)}_2]_{\ell-m+1,\ell+m'+1} }{i(\ell+1)} \frac{g_\YM^2}{16\pi^2} \frac{(-1)^{m'} }{x_3 y_3}\frac{\, _2F_1(\ell+1,\ell+1;2 \ell+2;-\xi^{-1} )}{\binom{2\ell+1}{\ell+1}\xi^{\ell+1}}\eqncom\nonumber
\end{align}
where $[t^{(2\ell+1)}_2]_{\ell-m+1,\ell-m'+1}=\frac{1}{2i} \bigl( \sqrt{(\ell+m)(\ell-m')}  \delta_{m',m-1}  - \sqrt{(\ell+m')(\ell-m)}\delta_{m',m+1} \bigr)$ denotes the respective matrix element of $t_2$ in the $(2\ell+1)$-dimensional irreducible representation.

\paragraph{Vacua}

There are three different types of two-point functions corresponding to protected  $\mathcal{N}=4$ SYM operators built from identical complex fields, namely
\begin{align}
&\langle \tr Z^{J_1}(x) \tr Z^{J_2}(y)\rangle,
&&\langle \tr Z^{J_1}(x) \tr \bar{Z}^{J_2}(y)\rangle
&&\langle \tr Z^{J_1}(x) \tr X^{J_2}(y)\rangle\eqndot
\end{align}
Let us first spell out the derivation of the two-point functions involving the BMN vacuum $\tr Z^{J_i}$ and its conjugate
\begin{align}
 \langle \tr Z^{J_1}(x) \tr \bar{Z}^{J_2}(y) \rangle_{\spec} & =J_1 J_2 \tr((\Zc)^{J_1-1} \contraction{}{\Zbar}{)(x)\tr(}{\Zbar}Z)(x)\tr(\Zbar (\Zbarc)^{J_2-1})(y)\\
 &=\frac{J_1J_2(-1)^{J_1+J_2}}{x_3^{J_1-1}y_3^{J_2-1}} \tr(t_3^{J_1-1}\hat{Y}_{\ell}^m)\tr(t_3^{J_2-1}\hat{Y}_{\ell'}^{m'}) \langle (Z)_{\ell m}(x)(\bar Z)_{\ell' m'}(y)\nonumber\rangle\eqndot
\end{align}
The contractions can be performed using~(\ref{eq: Contractions}) and 
the occurring traces can be calculated using the identities in appendix \ref{App:Fuzzy}.
We find
\begin{equation}
\label{eq: trZJ trZbarJ}
 \begin{aligned}
\langle \tr Z^{J_1} \tr \bar{Z}^{J_2} \rangle_{\spec} & =
\frac{g_\YM^2}{16\pi^2}
\frac{J_1}{x_3^{J_1}} \frac{J_2}{y_3^{J_2}}
\sum_{\ell=0}^{\infty}
\frac{\alpha^{J_1-1}_{\ell}\alpha^{J_2-1}_{\ell}}{\binom{2\ell+1}{\ell+1} }\frac{\, _2F_1(\ell,\ell+1;2\ell+2;-\xi^{-1} )}{\xi^{\ell+1}}\eqncom
\end{aligned}
\end{equation}
where we have used that $\tr(t_3^{J_i-1}\hat{Y}_{\ell}^m)=\alpha^{J_i-1}_{\ell}\delta^{m0}$ with $\alpha^{J_i-1}_{\ell}$ defined in \eqref{eq: definition of alpha}.
We have also dropped the sign as $\alpha^{J_1-1}_{\ell}\alpha^{J_2-1}_{\ell}$ vanishes unless $J_1+J_2$ is even.
Note that the sum is in fact finite, being restricted to $\ell<\min(k,J_1,J_2)$, due to the properties of  $\alpha^{J_i-1}_{\ell}$.
Further note that the $k$ dependence enters via $\alpha^{J_i-1}_{\ell}$, which depends on
\begin{equation}
\label{eq: definition curly B}
 \mathcal{B}_m = \frac{B_m(\tfrac{1-k}{2})}{m}\eqncom
\end{equation}
with $B_m$ being the Bernoulli polynomial of degree $m$.
Similarly,
\begin{equation}
\langle \tr Z^{J_1} \tr Z^{J_2} \rangle_{\spec} =
\frac{\xi }{\xi +1}\langle \tr Z^{J_1} \tr \bar{Z}^{J_2} \rangle_{\spec} \eqndot \label{2pt-unbarred}
\end{equation}
The final two-point function requires a bit more work. We need 
\begin{align}
 \langle \tr Z^{J_1}(x) \tr X^{J_2}(y) \rangle_{\spec} & =J_1 J_2 \tr((\Zc)^{J_1-1} \contraction{}{X}{)(x)\tr(}{X}Z)(x)\tr(X (X^{\mathrm{cl}})^{J_2-1})(y)\\
 &=\frac{J_1}{x_3^{J_1-1}} \frac{J_2}{y_3^{J_2-1}}\tr(t_3^{J_1-1}\hat{Y}_{\ell}^m)\tr(t_1^{J_2-1}\hat{Y}_{\ell'}^{m'}) \langle (Z)_{\ell m}(x)(X)_{\ell' m'}(y)\nonumber\rangle\\
 &=\frac{J_1}{x_3^{J_1-1}} \frac{J_2}{y_3^{J_2-1}}\tr(t_3^{J_1-1}\hat{Y}_{\ell}^0)\tr(t_1^{J_2-1}\hat{Y}_{\ell}^{\pm1}) \langle (Z)_{\ell 0}(x)(X)_{\ell \pm1}(y)\nonumber\rangle \eqndot
\end{align}
Then again via the results from appendix \ref{App:Fuzzy}, we arrive at
\begin{align}
\label{eq: trZJ tr XJ}
\langle \tr Z^{J_1} \tr X^{J_2} \rangle_{\spec} & =
 \frac{g_\YM^2}{16\pi^2}
\frac{J_1}{x_3^{J_1}} \frac{J_2}{y_3^{J_2}} 
\sum_{\ell=1}^\infty
\frac{i^{\ell+1}}{2^\ell}
\frac{\dbinom{\ell}{\frac{\ell-1}{2}} \alpha_\ell^{J_1-1}\alpha_\ell^{J_2-1}}{\binom{2\ell+1}{\ell+1}} 
\frac{{}_2F_1(\ell+1,\ell+1;2\ell+2;-\xi^{-1})}{\xi^{\ell+1}}
\eqndot
\end{align}

\paragraph{Vacuum and descendant}

In what follows, we will also be interested in the connected contribution to the two-point function $\langle \tr Z^{J_1} \tr X^{J_2-1} \bar{Z} \rangle$. This is straightforwardly worked out to be
\begin{align}
\langle \tr Z^{J_1} \tr X^{{J_2}-1} \bar{Z} \rangle_\spec = \frac{{J_1}}{x_3^{{J_1}-1} y_3^{{J_2}-1}}
\tr ( t_3^{{J_1}-1} \hat{Y}^m_{\ell})
\Bigl[& \langle Z_{\ell,m} \bar{Z}_{\ell',m'}\rangle   \tr (t_1^{{J_2}-1} \hat{Y}^{m'}_{\ell'}) \\&+
\sum_{p=0}^{{J_2}-2}\langle Z_{\ell,m} X_{\ell',m'}\rangle   \tr (t_1^{{J_2}-2-p} t_3 t_1^p \hat{Y}^{m'}_{\ell'})\nonumber
\Bigr]\eqncom 
\end{align}
which can be rewritten, using the explicit form of the propagators and the trace factors as evaluated in appendix  \ref{App:Fuzzy}, to the following form
\begin{align}\label{eq:propZXZ}
&\langle \tr Z^{{J_1}} \tr X^{{J_2}-1} \bar Z \rangle_{\spec} = \frac{g_\YM^2}{16\pi^2}\frac{J_1}{x_3^{J_1} y_3^{J_2}} \\
&\quad \sum_{\ell=0}^\infty 
\frac{i^\ell}{2^\ell} 
\frac{\binom{\ell}{\ell/2} \alpha_\ell^{{J_1}-1}\alpha_\ell^{{J_2}-1}}{\binom{2\ell+1}{\ell+1} \xi^{\ell+1}} 
\Big[ {}_2F_1(\ell,\ell+1;2\ell+2;-\xi^{-1}) + \ell \,{}_2F_1(\ell+1,\ell+1;2\ell+2;-\xi^{-1}) \Bigr]\eqndot
\nonumber
\end{align}
\vspace*{0.05cm}

\paragraph{Large $k$}

The fact that in the defect set-up one has an extra tunable parameter $k$ makes it possible to consider the following double-scaling limit~\cite{Nagasaki:2011ue,Nagasaki:2012re}, which allows
for a perturbative comparison of  string- and gauge-theory results:
\begin{equation}\label{dsl}
\lambda\rightarrow \infty\,, \hspace{0.5cm} k\rightarrow \infty\,, \hspace{0.5cm} \frac{\lambda}{k^2} \hspace{0.35cm}\mbox{finite}\,,
\end{equation}
where   $\lambda=g_\YM^2N$ is the 't Hooft coupling.
In this limit, assuming $\lambda/k^2$ to be small and assuming the field-theory observables to organise into a power series
in $\lambda/k^2$, one can treat field theory perturbatively 
while at the same time treating string theory in a supergravity approximation as justified by taking $\lambda\rightarrow \infty$. This strategy
has proved successful in the case of one-point functions where a precise match between field theory and string theory has been found 
for the one-point function of the BMN vacuum state both at 
leading~\cite{Nagasaki:2012re} and at next-to-leading order in the double-scaling parameter~\cite{Buhl-Mortensen:2016pxs,Buhl-Mortensen:2016jqo}. 
Moreover, field-theory considerations 
suggest an all-loop asymptotic formula for one-point functions of the SU(2) sector which in the case of the BMN vacuum state
agrees with the string-theory prediction of~\cite{Nagasaki:2012re} to all orders in the double-scaling parameter~\cite{Buhl-Mortensen:2017ind}.

 In the case of two-point functions, there does not at the moment exist any string-theory prediction but, obviously, it would  be very interesting to
derive one. In order to prepare for a future comparison with a string-theory calculation in the  double-scaling limit~(\ref{dsl}), 
we here present the  $k\to\infty$ limit of the two-point functions above.
Considering $J_1, J_2\gg1$ but finite, we can perform the sums over $\ell$ by considering \eqref{eq:alphaK} and find the following leading $k$ behaviour
\begin{align}
\langle \tr Z^{J_1} \tr \bar{Z}^{J_2} \rangle
&= \frac{\lambda}{16\pi^2}\frac{1}{N}\left(\frac{k}{2}\right)^{J_1+J_2-1}
\frac{1}{x_3^{J_1}y_3^{J_2}}\frac{2 \xi +1}{(\xi +1) \xi ^2}\eqncom\\
\langle \tr Z^{J_1} \tr Z^{J_2} \rangle
&= \frac{\lambda}{16\pi^2}\frac{1}{N}\left(\frac{k}{2}\right)^{J_1+J_2-1}
\frac{1}{x_3^{J_1}y_3^{J_2}}\frac{2 \xi +1}{(\xi +1)^2 \xi}\eqncom\\
\langle \tr Z^{J_1} \tr X^{J_2} \rangle
&= -\frac{\lambda}{16\pi^2}\frac{1}{N}\left(\frac{k}{2}\right)^{J_1+J_2-1}
\frac{1}{x_3^{J_1}y_3^{J_2}}\frac{4}{(2\xi +1)^2 }\eqndot
\end{align}
We notice that unlike two-point functions in pure ${\cal N}=4$ SYM theory the present two-point functions carry a factor $\frac{1}{N}$ in the 't Hooft expansion. This complies nicely with the string-theory picture where the computation of the dual object would amount
to the computation of a three-point function with two legs ending at the $AdS_5$ boundary (at the insertion points of the field-theory operator) and one
leg ending on the D5-brane in the interior of $AdS_5$, a computation which would necessitate the introduction of a string vertex.  
A successful strategy for this type of computations, at least in the heavy-heavy-light case,  has been 
developed in~\cite{ Zarembo:2010rr,Janik:2010gc} and it would be very interesting to  implement it in the present set-up.

\section{Mining for conformal data}
\label{sec:OPE}

In this section, we will use the operator expansions from section \ref{sec:OPEandBOE} to extract conformal data from our bulk-bulk two-point functions. First, we use the bulk OPE to obtain information on one-point functions and structure constants. Second, we will use the BOE to constrain the bulk-to-boundary couplings.

\subsection{Bulk operator product expansion}

Let us first focus on to the bulk OPE for the two-point function, cf.\ \eqref{eq: OPE for two-point}. We will consider two cases, namely those involving only the BPS operators built from identical fields such as $\tr (Z^L)$ and the special case involving also $\tr( \bar{Z}X^{L-1})$.

\subsubsection{BPS operators}

Let us consider the two-point function $\langle\tr Z^{J_1} \tr Z^{J_2}\rangle$. The OPE \eqref{eq: OPE for two-point} uses the full two-point functions, which includes the non-connected diagrams. In particular, 
\begin{equation}
\label{eq:FZZ}
 \begin{aligned}
f(\xi)&= J_1 J_2 \frac{g_\YM^2}{16\pi^2}2^{J_1+J_2}
\sum_{\ell=0}^{\infty}
\frac{\alpha^{J_1-1}_{\ell}\alpha^{J_2-1}_{\ell}}{\binom{2\ell+1}{\ell+1} }\frac{\, _2F_1(\ell,\ell+1;2\ell+2;-\xi^{-1} )}{ \xi^{\ell}(\xi+1)}\\
&\phaneq +a_{\tr Z^{J_1}} a_{\tr Z^{J_2}}+O(g_\YM^{4})\eqncom
\end{aligned}
\end{equation}
where the 
one-point functions to one-loop order for even $J$ 
are given by \cite{Buhl-Mortensen:2016pxs,Buhl-Mortensen:2016jqo,Buhl-Mortensen:2017ind}\footnote{In \cite{Buhl-Mortensen:2016jqo} non-supersymmetric boundary conditions where
employed for the $\ell = 0$ modes (cf.~section \ref{sec:action}) leading to a 
different finite-$N$ correction (see eq.~(6.22) of
 \cite{Buhl-Mortensen:2016jqo}). In the planar limit,
the $\ell = 0$ modes are subleading, and \eqref{eq:atrJ non-planar} agrees with
the results of \cite{Buhl-Mortensen:2016pxs,Buhl-Mortensen:2016jqo,Buhl-Mortensen:2017ind}.}
\begin{multline}
  a_{\tr Z^{J}} = -2^{J+1}\biggl[\mathcal{B}_{J+1} + \frac{g_\YM^2 J}{16\pi^2}
    \biggl(\mathcal{B}_{J-1}\left[N-k+\frac{J-1}{2}\right]\\
    -\sum_{i=0}^{\lfloor\frac{k-2}{2}\rfloor} (H_{k-i-1} - H_i) \left[\frac{k-2i-1}{2}\right]^{J-1}\biggr)
 +O(g_\YM^{4})\biggr]
  \label{eq:atrJ non-planar}
\eqncom
\end{multline}
with $H_n = \sum_{i=1}^n i^{-1}$  the harmonic numbers and $\calB$ 
defined in \eqref{eq: definition curly B}. For odd $J$, the one-loop correction to the one-point function vanishes.
Let us now study \eqref{eq: OPE for two-point} order by order.

\paragraph{Tree level}

Let us first consider the OPE at leading order in $g_\YM$. At this order,  \eqref{eq: OPE for two-point} reduces to
\begin{align}\label{eq:OPEg0}
2^{J_1+J_2+2} \xi^\frac{J_1+J_2}{2} \mathcal{B}_{J_1+1}\mathcal{B}_{J_2+1} = \sum_k \lambda^{(0)}_{\tr Z^{J_1}\tr Z^{J_2}}{}^{\mathcal{O}_k} a^{(0)}_{k} F_{\text{bulk}}(\Delta^{(0)}_k,
\delta J,\xi) \eqncom
\end{align}
where $ ^{(0)}$ stands for the leading order in $g_\YM$. In order to compare the left- and right-hand sides, we need the following useful identity
\begin{align}\label{eq:XIviaBulk} 
\xi^{n+1} = \sum_{m=0}^{\infty} \frac{\dbinom{m+n-\frac{\delta J}{2}}{m}\dbinom{m+n+\frac{\delta J}{2}}{m}}{\binom{2m+2n-1}{m}} F_{\text{bulk}}(2m+2n,\delta J,\xi)\eqncom
\end{align}
which holds for any $n,\delta J$. Inserting this relation in the OPE expansion \eqref{eq:OPEg0}, we obtain
\begin{multline}
\sum_{m=0}^{\infty} \frac{ \binom{m+J_1-1}{m}\binom{m+J_2-1}{m}  \mathcal{B}_{J_1+1}\mathcal{B}_{J_2+1}}{ \binom{2m+J_1+J_2-3}{m}} F_{\text{bulk}}(2m+J,\delta J,\xi)
\\=
 \sum_k\frac{ \lambda^{(0)}_{\tr Z^{J_1}\tr Z^{J_2}}{}^{\cO_k} a^{(0)}_{k}}{2^{J+2}} F_{\text{bulk}}(\Delta^{(0)}_k,\delta J,\xi)\eqncom
\end{multline}
where $J=J_1+J_2$ and $\delta J = J_1-J_2$.
Now, we can compare coefficients in the above sums. In particular, let us again group the sum on the right-hand side according to conformal dimension, i.e.\ we
write 
$\sum_k = \sum_{\Delta}\sum_{i: \Delta_i=\Delta}$: %
\begin{align}\label{eq:TreeOPEbulk}
 \binom{\Delta-3}{\frac{\Delta-J}{2}}\sum_{i\,:\,\Delta_i=\Delta}\lambda^{(0)}_{\tr Z^{J_1}\tr Z^{J_2}}{}^{\cO_{i}} a^{(0)}_{{i}}
=
 \binom{\frac{\Delta+\delta J}{2}-1}{\frac{\Delta-J}{2}}\binom{\frac{\Delta-\delta J}{2}-1}{\frac{\Delta-J}{2}} 2^{J+2}  \mathcal{B}_{J_1+1}\mathcal{B}_{J_2+1} \eqndot
\end{align}
In particular, if there is only one state propagating with a certain conformal dimension, the above relation fixes the product of the one-point function and the structure constant. 
Note that there is a non-trivial $k$ dependence in the above equation. In particular, the structure constant $\lambda$ does not depend on $k$, but the one-point function $a$ does. We will later exploit this fact to explicitly compute structure constants for BMN operators.

\paragraph{One-loop}
Consider the two-point function of two protected operators. The OPE side of the equation \eqref{eq: OPE for two-point} can be expanded to next-to-leading order in the coupling constant by writing $\lambda,a,\Delta$ as powers series in $g_\YM$
\begin{align}
&\lambda_{ij}{}^{k} = \sum_n \left(\lambda_{i j}{}^{k}\right)^{(n)}\left[\frac{g^2_{\YM}}{16\pi^2}\right]^n,
&&a_{k}= \sum_n a^{(n)}_{k}\left[\frac{g^2_{\YM}}{16\pi^2}\right]^n,
&&\Delta= \sum_n \Delta^{(n)}\left[\frac{g^2_{\YM}}{16\pi^2}\right]^n\eqndot
\end{align}
Here, we are assuming that the operators are normalised that they are containing no powers of $g_\YM$ at leading order.
Thus, if we expand \eqref{eq: OPE for two-point} up to order $g_\YM^2$, we find 
\begin{equation}\label{eq:OPEBulkg}
 \begin{aligned}
\sum_k \lambda_{i j}{}^{k} a_{k} F_{\text{bulk}}(\Delta_k,\delta\Delta,\xi) &= 
\sum_k \left(\lambda_{i j}{}^{k}\right)^{(0)} a^{(0)}_{k} F_{\text{bulk}}(\Delta^{(0)}_k,\delta\Delta,\xi) \\&\phaneq+
\frac{g^2_{\YM}}{16\pi^2} 
\sum_k
 \Big[
\left(\lambda_{i j}{}^{k}\right)^{(1)} a^{(0)}_{k}+\left(\lambda_{i j}{}^{k}\right)^{(0)} a^{(1)}_{k}\Big]F_{\text{bulk}}(\Delta^{(0)}_k,\delta\Delta,\xi) \\&\phaneq+\frac{g^2_{\YM}}{16\pi^2} 
\sum_k  \left(\lambda_{ij}{}^{k}\right)^{(0)} a^{(0)}_{k} \Delta^{(1)}_k F'_{\text{bulk}}(\Delta^{(0)}_k,\delta\Delta,\xi)+O(g_\YM^4)\eqncom
\end{aligned}
\end{equation}
where $F'_{\text{bulk}}(\Delta,\delta\Delta,\xi) = \partial_{\Delta} F_{\text{bulk}}(\Delta,\delta\Delta,\xi)$.
In order to compare our two-point function with the above expansion of the conformal block, we write both sides as a power series in $\xi$. By definition, we have
\begin{align}
F_{\text{bulk}}(\Delta,\delta\Delta,\xi)=
\xi^{\frac{\Delta}{2}}\sum_{n=0}^\infty \frac{(\frac{\Delta+\delta \Delta}{2})_n(\frac{\Delta-\delta \Delta}{2})_n}{(\Delta-1)_n n!}(-\xi)^n\eqncom
\end{align}
where $(\dots)_n$ denotes the Pochhammer symbol.
Thus,
\begin{align}
&F^\prime_{\text{bulk}}(\Delta,\delta\Delta,\xi) =
\half\Big[2\Psi(\Delta-1)-
\Psi({\textstyle \frac{\Delta-\delta \Delta}{2}})-\Psi({\textstyle \frac{\Delta+\delta \Delta}{2}})
+\log\xi
\Big]F_{\text{bulk}}(\Delta,\delta\Delta,\xi) \\
&  -\xi^{\frac{\Delta}{2}} \sum_n \frac{(\frac{\Delta+\delta \Delta}{2})_n(\frac{\Delta-\delta \Delta}{2})_n}{(\Delta-1)_n n!}(-\xi)^n\Big[
\Psi(\Delta-1+n)-
\frac{\Psi({\textstyle \frac{\Delta-\delta \Delta}{2}}+n)+\Psi({\textstyle \frac{\Delta+\delta \Delta}{2}}+n)}{2}
\Big]
\eqncom\nonumber
\end{align}
where $\Psi$ is the Euler digamma function. 

\paragraph{Example}
Let us now try to work out the OPE for the case $\cO_1=\cO_2=\tr Z^2$. At tree-level, the OPE \eqref{eq:TreeOPEbulk} implies
\begin{align}
 \binom{\Delta-3}{\frac{\Delta}{2}-2}\sum_{i\,:\,\Delta_i=\Delta}\lambda^{(0)}_{\tr Z^{2}\tr Z^{2}}{}^{\cO_{i}} a^{(0)}_{{i}}
=
\frac{( \Delta-2)^2}{36} k^2(k^2-1)^2\,.
\end{align}
For $\Delta=4$, only the states $\tr Z^4$, $\tr Z ^3\tr Z $ and $\tr Z ^2\tr Z ^2$ are summed over. However, since the structure constants do not depend on $k$, while $a^{(0)}_{\tr Z ^2\tr Z ^2}=a^{(0)}_{\tr {Z}^2}a^{(0)}_{\tr {Z}^2}\sim\mathcal{B}_3^2$, $a^{(0)}_{\tr Z ^3\tr {Z}}=0$
and $ a^{(0)}_{\tr Z ^4} \sim \mathcal{B}_5$, we find that the above OPE can only be satisfied for 
\begin{align}
&\lambda^{(0)}_{\tr Z^{2}\tr Z^{2}}{}^{\tr Z ^4} = 0,
&&
&\lambda^{(0)}_{\tr Z^{2}\tr Z^{2}}{}^{\tr Z ^2\tr Z ^2} = 1\eqndot
\end{align}
This agrees with an explicit computation of the structure constant. Actually, it is not hard to see that at tree-level the structure constants 
take the following simple form
$\lambda^{(0)}_{\cO_i\cO_j}{}^{\cO_k} = \delta_{\cO_i \cO_j,\cO_k}$ when there are no
contractions possible between $\cO_i$ and $\cO_j$.

Let us then continue to one-loop. It is easy to see that for $J_1=J_2=2$, \eqref{eq:FZZ} simplifies to
\begin{align}
f_{\spec}(\xi)=\frac{g_\YM^2}{16\pi^2} \frac{16k(k^2-1)}{3} \Big[1+ \xi+ 2\xi\log\xi+\sum_{n=1}^{\infty} \frac{(-1)^n(n+2)}{n} \xi^{n+1}\Big]\eqncom
\end{align}
where we used that $(\alpha^1_1)^2 =\frac{k(k^2-1)}{12}$.
Then, the OPE implies
\begin{align}\label{eq:OneLoopOPEBPS}
&\frac{4}{3}k^2 (k^2-1)\Big[2N-k\Big]\xi^2 +\frac{16k(k^2-1)}{3} \Big[\xi^2+ \xi^3+ 2\xi^3\log\xi+\sum_{n=1}^{\infty} \frac{(-1)^n(n+2)}{n} \xi^{n+3}\Big] =\nonumber \\
&
\sum_k
\xi^{\frac{\Delta_k}{2}}\sum_{n=0}^\infty \frac{(\frac{\Delta_k}{2})_n(\frac{\Delta_k}{2})_n}{(\Delta_k-1)_n }\frac{(-\xi)^n }{n!}
\Bigg[
\left(\lambda_{i j}{}^{k}\right)^{(1)} a^{(0)}_{k}+\left(\lambda_{i j}{}^{k}\right)^{(0)} a^{(1)}_{k}
+ \\
& \left(\lambda_{i j}{}^{k}\right)^{(0)} a^{(0)}_{k} \Delta^{(1)}_k 
\Big\{
\Psi(\Delta-1)-\Psi(\Delta-1+n)-\Psi(\textstyle{\frac{\Delta}{2}})+\Psi(\textstyle{\frac{\Delta}{2}}+n)+\half\log\xi
\Big\}
\Bigg]\nonumber\eqndot
\end{align}
Again, let us restrict to the case $\Delta=4$. We can compare the terms proportional to $\xi^2$ and to $\xi^2\log\xi$ in \eqref{eq:OneLoopOPEBPS}. From $\xi^2\log\xi$, we find
\begin{align}
&
\lambda^{(0)}_{\tr Z^2 \tr Z^2}{}^{\tr Z ^4} a^{(0)}_{\tr Z ^4}\Delta^{(1)}_{\tr Z ^4}+
\lambda^{(0)}_{\tr Z^2 \tr Z^2}{}^{\tr Z ^2\tr Z ^2} a^{(0)}_{\tr Z ^2\tr Z ^2}\Delta^{(1)}_{\tr Z ^2\tr Z ^2} =0\eqncom
\end{align}
which implies
\begin{align}
\Delta^{(1)}_{\tr Z ^2\tr Z ^2} =0\eqndot
\end{align}
In other words, $\tr Z ^2\tr Z ^2$ is a protected operator. Since both $\tr Z^2\tr Z ^2$ and $\tr Z ^4$ are protected, their structure constants do not 
receive loop corrections~\cite{Heslop:2001gp,Baggio:2012rr}
\begin{align}
\lambda^{(1)}_{\tr Z^2 \tr Z^2}{}^{\tr Z ^4}=\lambda^{(1)}_{\tr Z^2 \tr Z^2}{}^{\tr Z ^2 \tr Z ^2}=0\eqndot
\end{align}
This leaves us with the following contribution from the $\xi^2$ term
\begin{align}
 a^{(1)}_{\tr {Z}^2\tr {Z}^2}
 = \frac{8}{3}k (k^2-1)\Big[2Nk-k^2+2\Big] \eqncom
\end{align}
where we used that $ \lambda^{(0)}_{\tr Z^2 \tr Z^2}{}^{\tr {Z}^3 \tr {Z}}=0$. This can be directly checked by a calculation in the quantum field theoretic framework of \cite{Buhl-Mortensen:2016jqo}. Finally, from the two-point function $\langle\tr {Z}^3\tr {Z}\rangle$, we similarly obtain
\begin{align}
 a^{(1)}_{\tr {Z}^3\tr {Z}}  = 4 k(k^2-1)\eqndot
\end{align}
Thus from the OPE we are able to derive non-planar one-loop one-point functions of multi-trace operators.

\subsubsection{BMN operators}

In the case of the two-point function $\langle \tr Z^{J_1} \tr X^{{J_2}-1} \bar{Z} \rangle$, only single-trace operators in the SU(2) sector contribute in the OPE channel at leading $\xi$ and leading $N$. Moreover, the one-point functions have a distinct $k$ dependence, which allows us to extract certain structure constants exactly. Notice that one of our states has a single impurity such that the disconnected part of the two-point function vanishes. 

For concreteness, let us look at the simple example $J_1=J_2=3$ first.
In the limit $\xi\to0$, this two-point function \eqref{eq:propZXZ} behaves as
\begin{equation}
\label{eq: limit of example}
\langle \tr Z^{3} \tr X^{2} \bar{Z} \rangle_\spec =  \frac{g_\YM^2}{16 \pi^{2}}\frac{k(k^4-1)}{80}\frac{1}{\xi}\frac{1}{x_3^3 y_3^3} +O(\xi^0)\eqndot
\end{equation}
We know that only two operators can propagate in the OPE channel: $\mathcal{K}$, a superconformal descendant of the Konishi primary operator, and $S_-^2[\tr Z^4]$, the second SU(2) descendant of the BPS vacuum $\tr(Z^4)$, see appendix \ref{app: descendants}. In the planar limit, these operators diagonalise $M_{ij}$:
\begin{align}
\mathcal{K} &= 
\tr (Z^2 X^2)-\tr (Z X Z X)\eqncom \\
S_-^2[\tr Z^4]&= 
8\tr (Z^2 X^2)+4\tr (Z X Z X) \eqndot
\end{align}
Their one-point functions are 
\begin{align}
 \langle {\mathcal{K}}\rangle&= -
 \frac{k(1-k^2)}{24}\frac{1}{x_3^4}\eqncom\\
 \langle S_-^2[\tr{Z}^4]\rangle&=
 \frac{k(1-k^2)(7-3k^2)}{60}\frac{1}{x_3^4}\eqndot
\end{align}
Again from the bulk OPE, we can fix the two structure constants by the non-trivial $k$ dependence of the one-point functions:
\begin{align}
&\lambda_{\tr Z^3\tr X^2\bar{Z}}{}^{{\mathcal{K}}}=\frac{g_\YM^2}{4\pi^2}\eqncom
&&\lambda_{\tr Z^3\tr X^2\bar{Z}}{}^{S_-^2[\tr{Z}^4]}=\frac{g_\YM^2}{16\pi^2}\eqndot
\end{align}
From this, we can compute the structure constants of these operators in the planar limit. If we normalise all operators such that they have two-point functions which are normalised to unity far away from the defect, then
\begin{align}
&\bar{\lambda}_{\tr Z^3\tr X^2\bar{Z}\bar{\mathcal{K}}}=\frac{1}{N}+O\left(\frac{1}{N^2}\right)\eqncom
&&\bar{\lambda}_{\tr Z^3\tr X^2\bar{Z}S_-^2[\tr\bar{Z}^4]}=\frac{\sqrt{2}}{N}+O\left(\frac{1}{N^2}\right)\eqncom
\end{align}
where $\bar{\lambda}$ stands for the normalised structure constant. This can easily be verified by explicit computation.
 
In the more general case of arbitrary odd $J_1=J$ and $J_2=2$, the operators which can appear in the OPE channel are the 
BMN operators ${\cal O}_n^{J-1}$ with $n=0,1,2,\ldots, (J-1)/2$~\cite{Berenstein:2002jq}, see e.g.~\cite{Beisert:2003tq} for the precise normalisation:
\begin{align}
{\cal O}_n^{J-1}=\frac{1}{\sqrt{J}}\frac{1}{(\sqrt{2})^{\delta_{n,0}}}\left(\frac{4\pi^2}{g_\textup{YM}^2 N}\right)^{\frac{J+1}{2}}\Bigg\{&
\sum_{m=0}^{(J-3)/2} 2 \cos \left(\frac{\pi n(2m+1)}{J}\right) \tr (X Z^m X Z^{J-m-1}) \nonumber \\
& +\cos (\pi n) 
\tr (X Z^{\frac{J-1}{2} }X Z^{\frac{J-1}{2}}) \Bigg\}\eqncom
\end{align}
where the case $n=0$ corresponds to a descendant of the vacuum. We again work in the planar limit.
Next, we need to consider the two-point function. To leading order in the $\xi\to0$ limit, the two-point function \eqref{eq:propZXZ} reduces to
\begin{align}
\langle \tr Z^{J} \tr X^{2} \bar{Z} \rangle \to\frac{g_\YM^2}{16 \pi^{2}}J\left[ \mathcal{B}_{J+2} - \frac{k^2-1}{4}\mathcal{B}_{J} \right]\frac{1}{\xi}\frac{1}{x_3^J y_3^3}+O(\xi^0)\eqndot
\eqndot
\end{align}
Again, there are two types of states running in the OPE channel: the descendant of the vacuum and Bethe states with non-trivial momentum. The one-point function of the descendant follows directly from appendix \ref{app: descendants}:
\begin{align}
a_{{\cal O}_0^{J-1}} = -2^{J+1}\sqrt{\frac{2}{J}}\left(\frac{4\pi^2}{g^2_\YM N}\right)^{\frac{J+1}{2}}\,\mathcal{B}_{J+2}\eqndot
\end{align}
Note that ${\cal O}_n^{J-1}$ is normalised to have a unit-normalised two-point function far away from the defect, which affects its one-point function. 
The one-point functions of the Bethe states take the form \cite{Buhl-Mortensen:2015gfd}
\begin{align}
a_{{\cal O}_n^{J-1}} = 
2^{J+1}\left(\frac{4\pi^2}{g^2_\YM N}\right)^{\frac{J+1}{2}}\,\frac{1}{\sqrt{J}}\sqrt{\frac{u_n^2+\frac{1}{4}}{u_n^2}}\sum_{j=\frac{1-k}{2}}^{\frac{k-1}{2}} j^L \frac{u_n^2(u_n^2+\frac{k^2}{4})}{[u_n^2 + (j-\half)^2][u_n^2 + (j+\half)^2]}\eqncom
\end{align}
where $u_n = \half\cot \frac{\pi \, n}{J}$. 
From appendix \ref{app: one-point functions}, we can recast the structure constant side of the OPE as a sum over $\mathcal{B}_{m}$'s, which allows us to determine the structure constants $\lambda$ by comparing coefficients in front of the different Bernoulli polynomials. Moreover, we have can write the two-point function side of the OPE also in terms of Bernoulli polynomials via
\begin{align}
\frac{k^2-1}{4}\mathcal{B}_{L-1} = \frac{L+1}{L-1}\mathcal{B}_{L+1}-2\sum_{n=1}^{L/2} b_{L-2n+2}\frac{\Gamma(L-1)}{\Gamma(2n-2)\Gamma(L-2n+3)}\mathcal{B}_{2n-1}\eqncom
\end{align}
where $b_n$ is the $n$th Bernoulli number.

By then considering the coefficient in front of $\mathcal{B}_{J+2}$, we can immediately read off the structure constant of the descendant operator. More precisely, we have
\begin{align}
&-2 \mathcal{B}_{J+2} N = -\sqrt{2} \mathcal{B}_{J+2} \bar{\lambda}_{\tr Z^J\tr X^2\bar{Z}\,\bar{\cO}^{J-1}_n},
&&\Rightarrow
&& \bar{\lambda}_{\tr Z^J\tr X^2\bar{Z}\,\bar{\cO}^{J-1}_n} = \frac{\sqrt{2}}{ N}.
\end{align}
Again, the $\bar{\lambda}$ stands for the structure constants where all the operators are normalised to have unit two-point functions far away from thee defect. More generally, we obtain the structure constants
\begin{equation}
\bar{\lambda}_{\tr Z^J \tr X^2\bar{Z}\,\bar{\cO}_n^{J-1}} =\frac{1}{N}\frac{2 \cos (\frac{\pi n}{J})}{(\sqrt{2})^{\delta_{n,0}}}\eqndot
\end{equation}
The above structure constants can also be computed directly using the standard Wick contractions of  
${\cal N}=4$ SYM theory and the result obtained in this way  fully agrees with the above result.

\paragraph{More excitations} For more than two excitations, the situation changes. As can be seen from appendix \ref{app: one-point functions}, the one-point functions of an operator in the SU(2) sector with length $L$ and $M$ excitations are polynomials of degree $L-M+1$ in $k$. In particular, the number of constraints that the OPE imposes grows linearly with the length of the operators. However, it can be quickly seen that the number of Bethe states grows polynomially. For example, for four excitations, the number of Bethe states with paired rapidities grows quadratically in $L$. This means that the structure constants after a certain length will not be completely fixed from just the leading contribution to the connected two-point function. However, more constraints arise if one goes to subleading order and in this way the structure constants can always be derived. 

For some states, one can nevertheless fix the structure constant for any length from the leading order. Since the one-point functions have degree $L-M+1$, descendants have a different degree than Bethe states. In particular, the only state with a term $k^{L+1}$ is the $M$th descendant of the vacuum, which according to appendix \ref{app: descendants} has the following one-point function to leading order 
\begin{equation}
a_{S^M_-[\tr Z^L]} = -2^{L+1}\frac{M!(\frac{L}{2})!}{(\frac{M}{2})!(\frac{L-M}{2})!}\mathcal{B}_{L+1}
 \eqndot
\end{equation}
This allows us to compute the corresponding structure constant in the planar limit
\begin{align}
\bar{\lambda}_{ \tr Z^{L-M} \tr X^M\bar{Z}\, S_-^M[\tr\bar{Z}^L]} = \frac{1}{N}\sqrt{\frac{(L-M)! M!}{(L-2)!}}\eqndot
\end{align}

\subsection{Boundary operator expansion}
\label{sec:BOE}

Let us now move on to the BOE \eqref{eq: BOE for two-point}. We can use our bulk-bulk two-point functions \eqref{eq: trZJ trZbarJ}, \eqref{2pt-unbarred} and \eqref{eq: trZJ tr XJ} to find bulk-to-boundary couplings.

In order to equate our bulk-bulk two-point functions to the boundary expansion \eqref{eq: BOE for two-point}, we need to express them in terms of the boundary conformal blocks $F_{\text{bdy}}$ defined in \eqref{eq: boundary conformal block}. In fact, we have the following identities relating the hypergeometric functions in the two-point functions to $F_{\text{bdy}}$: 
\begin{align}
\frac{{}_2F_1(\Delta-1,\Delta;2\Delta;-\xi^{-1})}{\xi^\Delta} 
& = F_{\text{bdy}}(\Delta,\xi)+ \frac{1}{2}F_{\text{bdy}}(\Delta+1,\xi)+ \frac{1-\Delta^2}{\frac{1}{4}-\Delta^2}\frac{F_{\text{bdy}}(\Delta+2,\xi)}{16}\eqncom\nonumber\\
\frac{\xi}{\xi+1}\frac{{}_2F_1(\Delta-1,\Delta;2\Delta;-\xi^{-1})}{\xi^\Delta} 
& = F_{\text{bdy}}(\Delta,\xi)- \frac{1}{2}F_{\text{bdy}}(\Delta+1,\xi)+ \frac{1-\Delta^2}{\frac{1}{4}-\Delta^2}\frac{F_{\text{bdy}}(\Delta+2,\xi)}{16}\eqncom\nonumber\\
\frac{{}_2F_1(\Delta,\Delta;2\Delta;-\xi^{-1})}{\xi^\Delta} 
& = F_{\text{bdy}}(\Delta,\xi) + \frac{\Delta(\Delta+1)}{\frac{1}{4}-\Delta^2}\frac{F_{\text{bdy}}(\Delta+2,\xi)}{16}\eqndot
\end{align}
This allows us to immediately compare the two-point functions \eqref{eq: trZJ trZbarJ}, \eqref{2pt-unbarred} and \eqref{eq: trZJ tr XJ} to the boundary conformal block structure \eqref{eq: BOE for two-point}. In particular, we find
\begin{align}\label{eq:PropviaBDY}
\sum_i \mu_{\tr Z^{J_1}}{}^{\hat\cO_i}\mu_{\tr \bar{Z}^{J_2}\hat\cO_i}F_{\text{bdy}}(\Delta_i,\xi) &= 
\frac{g_\YM^2J_1 J_2 2^{J_1+J_2}}{16\pi^2} 
\sum_{\ell=0}^{\infty}
\frac{\alpha^{J_1-1}_{\ell}\alpha^{J_2-1}_{\ell}}{\binom{2\ell+1}{\ell+1} }\Bigg[F_{\text{bdy}}(\ell+1,\xi)\\
&\phaneq+\frac{1}{2}F_{\text{bdy}}(\ell+2,\xi)+\frac{1}{4}\frac{\ell(\ell+2)}{(2\ell+1)(2\ell+3)}F_{\text{bdy}}(\ell+3,\xi)\Bigg],\nonumber \\
\sum_i \mu_{\tr Z^{J_1}}{}^{\hat\cO_i}\mu_{\tr Z^{J_2}\hat\cO_i}F_{\text{bdy}}(\Delta_i,\xi) &= 
\frac{g_\YM^2J_1 J_22^{J_1+J_2}}{16\pi^2} 
\sum_{\ell=0}^{\infty}
\frac{\alpha^{J_1-1}_{\ell}\alpha^{J_2-1}_{\ell}}{\binom{2\ell+1}{\ell+1} }\Bigg[F_{\text{bdy}}(\ell+1,\xi)\\
&\phaneq-\frac{1}{2}F_{\text{bdy}}(\ell+2,\xi)+\frac{1}{4}\frac{\ell(\ell+2)}{(2\ell+1)(2\ell+3)}F_{\text{bdy}}(\ell+3,\xi)\Bigg] ,\nonumber \\
\sum_i \mu_{\tr Z^{J_1}}{}^{\hat\cO_i}\mu_{\tr X^{J_2}\hat\cO_i}F_{\text{bdy}}(\Delta_i,\xi) &= 
\frac{g_\YM^2J_1 J_22^{J_1+J_2}}{16\pi^2} 
\sum_{\ell=1}^{\infty}
\frac{i^{\ell+1}}{2^\ell}
\frac{\dbinom{\ell}{\frac{\ell-1}{2}} \alpha_\ell^{J_1-1}\alpha_\ell^{J_2-1}}{\binom{2\ell+1}{\ell+1}} \\
&\phaneq
\times\Bigg[F_{\text{bdy}}(\ell+1,\xi)+\frac{1}{4}\frac{(\ell+1)(\ell+2)}{1-4(\ell+1)^2}F_{\text{bdy}}(\ell+3,\xi)\Bigg] \nonumber.
\end{align}
Since the hypergeometric functions are independent, we can directly read off the spectrum and the bulk-to-boundary couplings $\mu$.
More precisely, we find that the spectrum running in the boundary channel has $\Delta=1,2,\ldots, \min(J_1,J_2,k)+2$.  
Let us group the sum on the boundary side according to conformal dimension $\sum_i = \sum_{\Delta}\sum_{i:\Delta_i=\Delta}$. Then, we find the following set of equations for the bulk-to-boundary couplings:
\begin{equation}\label{eq:BOPEdefect}
 \begin{aligned}
\sum_{i:\Delta_i=\Delta}
\frac{\mu_{\tr Z^{J_1}}{}^{\hat\cO_{i}}\mu_{\tr \bar{Z}^{J_2}\hat\cO_{i}}}{\frac{g_\YM^2}{16\pi^2} J_1 J_22^{J_1+J_2}}
&= \frac{\alpha^{J_1-1}_{\Delta-1}\alpha^{J_2-1}_{\Delta-1}}{\binom{2\Delta-1}{\Delta} } +
\frac{\alpha^{J_1-1}_{\Delta-2}\alpha^{J_2-1}_{\Delta-2}}{2\binom{2\Delta-3}{\Delta-1} }+
\frac{(\Delta-1)(\Delta-3)}{(2\Delta-3)(2\Delta-5)}\frac{\alpha^{J_1-1}_{\Delta-3}\alpha^{J_2-1}_{\Delta-3}}{4\binom{2\Delta-5}{\Delta-2} } \eqncom\\
\sum_{i:\Delta_i=\Delta}
\frac{\mu_{\tr Z^{J_1}}{}^{\hat\cO_{i}}\mu_{\tr Z^{J_2}\hat\cO_{i}}}{\frac{g_\YM^2}{16\pi^2} J_1 J_22^{J_1+J_2}}
&= \frac{\alpha^{J_1-1}_{\Delta-1}\alpha^{J_2-1}_{\Delta-1}}{\binom{2\Delta-1}{\Delta} } -
\frac{\alpha^{J_1-1}_{\Delta-2}\alpha^{J_2-1}_{\Delta-2}}{2\binom{2\Delta-3}{\Delta-1} }+
\frac{(\Delta-1)(\Delta-3)}{(2\Delta-3)(2\Delta-5)}\frac{\alpha^{J_1-1}_{\Delta-3}\alpha^{J_2-1}_{\Delta-3}}{4\binom{2\Delta-5}{\Delta-2} } \eqncom\\
\sum_{i:\Delta_i=\Delta}
\frac{\mu_{\tr Z^{J_1}}{}^{\hat\cO_{i}}\mu_{\tr X^{J_2}\hat\cO_{i}}}{\frac{g_\YM^2}{16\pi^2} J_1 J_22^{J_1+J_2}}
&= 
\frac{\dbinom{\Delta-1}{\frac{\Delta-2}{2}} \alpha_{\Delta-1}^{J_1-1}\alpha_{\Delta-1}^{J_2-1}}{i^{-\Delta}2^{\Delta-1}\binom{2\Delta-1}{\Delta}}
-
\frac{\dbinom{\Delta-3}{\frac{\Delta-4}{2}} \alpha_{\Delta-3}^{J_1-1}\alpha_{\Delta-3}^{J_2-1}}{i^{-\Delta}2^{\Delta-1}\binom{2\Delta-5}{\Delta-2}} 
\frac{(\Delta-2)(\Delta-1)}{1-4(\Delta-2)^2}
\,,
\end{aligned}
\end{equation}

where $\alpha^{J}_{-1}=\alpha^{J}_{-2}=0$ and the binomial $\dbinom{\Delta}{\frac{\Delta-1}{2}}$ is understood to vanish for $\Delta<1$. Notice furthermore that, since $\alpha^{J}_n=0$ if $J+n=\mathrm{odd}$, at most one or two terms in the above expression actually contributes.

Let us now consider $\Delta=2$ with $J_1=J_2=2$.
Two kinds of multiplets on the boundary can in principle contribute, cf.\ appendix \ref{app: boundary operators}; 
they transform as $(0,0)$ and $(2,0)$ of $\soc\times\soe$, respectively.
Thus, we find from \eqref{eq:BOPEdefect}
\begin{equation}
 \sum_{\alpha}(\mu_{\tr Z^{2}\hat{\mathcal O}_{[2,(0,0)],\alpha}})^2
 +\sum_{\beta}(\mu_{\tr Z^{2}\hat{\mathcal O}_{[2,(2,0)],0,\beta}})^2
 =\frac{g_\YM^2}{16\pi^2}\frac{4\times2^4(\alpha_1^1)^2}{3}\eqncom
\label{eq:BOPEdefect delta 2 part 1}
\end{equation}
and
\begin{multline}
 \sum_{\alpha}\mu_{\tr Z^{2}\hat{\mathcal O}_{[2,(0,0)],\alpha}}\mu_{\tr X^{2}\hat{\mathcal O}_{[2,(0,0)],\alpha}}
 +\sum_{\beta}\mu_{\tr Z^{2}\hat{\mathcal O}_{[2,(2,0)],0,\beta}}\mu_{\tr X^{2}\hat{\mathcal O}_{[2,(2,0)],0,\beta}}\\
 =-\frac{1}{2}\frac{g_\YM^2}{16\pi^2}\frac{4 \times2^4(\alpha_1^1)^2}{3}\eqncom
\label{eq:BOPEdefect delta 2 part 2}
\end{multline}
We have chosen a real basis of boundary operators which is furthermore
unit-normalised with respect to their respective boundary-boundary two-point functions. It follows that
we can freely raise and lower the second index on $\mu$.

As shown in appendix \ref{app: boundary operators} we have 
\begin{equation}
  \mu_{\tr X^{2}\hat{\mathcal O}_{[2,(0,0)],\alpha}} = \mu_{\tr Z^{2}\hat{\mathcal O}_{[2,(0,0)],\alpha}}\eqncom\qquad
  \mu_{\tr X^{2}\hat{\mathcal O}_{[2,(2,0)],0,\beta}} = -\frac{1}{2} \mu_{\tr Z^{2}\hat{\mathcal O}_{[2,(2,0)],0,\beta}}\eqndot
\end{equation}
Combining this with \eqref{eq:BOPEdefect delta 2 part 1} and \eqref{eq:BOPEdefect delta 2 part 2} yields
\begin{equation}
 \sum_{\alpha}(\mu_{\tr Z^{2}\hat{\mathcal O}_{[2,(0,0)],\alpha}})^2 = 0 \eqncom \qquad
 \sum_{\beta}(\mu_{\tr Z^{2}\hat{\mathcal O}_{[2,(2,0)],0,\beta}})^2 = \frac{g_\YM^2}{\pi^2}\frac{4(\alpha_1^1)^2}{3}\eqndot
  \label{eq:BOPEdefect delta 2 part 3}
\end{equation}
By $\soe$ symmetry we have $\mu_{\tr Z^2\hat{\mathcal O}_i} = \mu_{\tr\bar{Z}^2\hat{\mathcal O}_i}$
for any $\hat{\mathcal O}_i$ which is a singlet under $\soe$. 
But then $\mu_{\tr Z^{2}\hat{\mathcal O}_{[2,(0,0)],\alpha}}$ is real, and we 
conclude from \eqref{eq:BOPEdefect delta 2 part 3} that
\begin{equation}
  \mu_{\tr Z^{2}\hat{\mathcal O}_{[2,(0,0)],\alpha}} = 0\eqncom 
\end{equation}
for all $\alpha$. As we see, the dimension-two $R$-singlets decouple to leading order in $g_\YM$.
It is tempting to speculate that there is some symmetry underlying this result. We leave further
investigation of this to future work.

\section{Conclusions \& Outlook\label{sec:Conclusion}}

Numerous novel types of multi-point correlation functions appear when defects are introduced in a conformal field theory.  With the present paper, we have initiated the calculation of such correlation functions in the case of a defect version of ${\cal N}=4$ SYM theory dual to the D5-D3 probe-brane system with background gauge field flux. Apart from being interesting in their own
right, these correlation functions have the prospect of serving as input to the conformal bootstrap program both for 
${\cal N}=4$ SYM theory itself~\cite{Beem:2013qxa,Beem:2016wfs} and for its defect version~\cite{Liendo:2012hy,Gliozzi:2015qsa,Billo:2016cpy,Liendo:2016ymz}. We have illustrated this by using the knowledge of one- and two-point functions of the dCFT
to extract  structure constants of ${\cal N}=4$ SYM theory from the bulk OPE  and bulk-to-boundary couplings
from the BOE. This type of exploitation of the OPE and the associated crossing relations has also previously proven 
very efficient in accessing information about higher-loop correlation functions, e.g.\ the five-loop correction to the anomalous
dimension of the Konishi operator~\cite{Eden:2012fe}.  

In order to make further progress on the present dCFT, it is essential to derive the explicit form of the  3D defect action.  So far, this has only been
accomplished for the simpler case of $k=0$~\cite{DeWolfe:2001pq}. We have already presented the complete spectrum of boundary operators in the case $k\neq 0$
in appendix~\ref{app: boundary operators}.  The task is now to constrain the possible interaction terms involving these fields invoking the OSp$(4|4)$ symmetry of
the system.
 
There exists a couple of somewhat related defect versions of ${\cal N}=4$ SYM theory which in the string-theory language are generated by
introducing  a D7 probe-brane with geometry either $AdS_4\times S^2 \times S^2$ or $AdS_4\times S^4$ and correspondingly with
background gauge field flux on either $S^2\times S^2$ or on $S^4$. These defect CFTs, for which supersymmetry is completely broken, have so far only been considered at tree level~\cite{Kristjansen:2012tn,deLeeuw:2016ofj}. It would be interesting to set up the perturbative program for these theories as well and in particular to investigate to which extent the absence of supersymmetry complicates or changes the present analysis. Another dCFT more closely related to the one considered
in this paper is  ${\cal N}=4$ SYM theory with a line defect which has exactly the same symmetry group as the present dCFT, namely OSp$(4|4)$; see for instance~\cite{Liendo:2016ymz}. As pointed out in~\cite{Liendo:2016ymz}, the part of the analysis pertaining to the boundary conformal bootstrap equations 
can be carried over to the case of the line defect. Developing the perturbative analysis of the corresponding dCFT would  be interesting as well.

As mentioned earlier, it has previously been possible to match  one-point functions calculated in the defect field theory with one-point functions
calculated in the dual string theory  in a certain double-scaling limit both at the classical~\cite{Nagasaki:2012re,Buhl-Mortensen:2015gfd} and at the quantum level~\cite{Buhl-Mortensen:2016pxs,Buhl-Mortensen:2016jqo}.  
It would likewise be very interesting to  perform a calculation of two-point functions in the string-theory language and to check the agreement with the field-theory prediction.  Such a calculation would amount to evaluating a three-point function of classical strings in the spirit
of~\cite{ Zarembo:2010rr,Janik:2010gc,Klose:2011rm}
with two strings ending at the $AdS_5$ boundary (at the insertion points of the two-point functions) and one ending on the D5-brane in the interior
of $AdS_5$. Regarding correlation functions, the understanding of the dCFT is currently more complete than that of the corresponding probe-brane system and progress on the string-theory side would be very important for the further exploration of AdS/dCFT.

\begin{acknowledgments}
We thank I.\ Buhl-Mortensen, P.\ Liendo, G. Semenoff and K.\ Zarembo  for useful discussions.
The authors were  supported  in  part  by  FNU  through grants number DFF-1323-00082  and
 DFF-4002-00037. M.d.L.\ further acknowledges support by the ERC advanced grant 291092.
\end{acknowledgments}

\appendix

\section{One-point functions of descendants\label{App:Descendants}}
\label{app: descendants}

In \cite{deLeeuw:2015hxa,Buhl-Mortensen:2015gfd}, the tree-level one-point functions of primary operators in the SU(2) sector have been calculated via integrability as normalised overlaps of Bethe eigenstates with a matrix product state:
\begin{equation}
a_{\mathcal{O}_{L,M,\{u\}}}=2^{L}\left(\frac{4\pi^2}{g_{\YM}^2N}\right)^{\frac{L}{2}}\frac{C_k}{\sqrt{L}}  \eqncom \qquad C_k=\frac{\langle\mbox{MPS} |\{u_j\}\rangle}{\langle\{u_j\} |\{u_j\}\rangle^{1/2}}\eqndot 
\end{equation}
Notice that in this definition the operators $\mathcal{O}_{L,M,\{u\}}$ are primary and normalised such that $ M_{ij} = \delta_{ij}$ in the planar limit. Extra care needs to be taken in the  case of one-point functions for descendant operators, which are the topic of this appendix.

Descendant states can be obtained from the highest weight Bethe eigenstates by sending some of the rapidities to infinity. This process is most cleanly described using the coordinate Bethe ansatz, 
where it holds that~\cite{Escobedo:2010xs}
 \begin{align}
 \lim_{u_k\rightarrow \infty} |\{u_j\}\rangle^{\mbox{\footnotesize co}}=S_{-}|\{u_j\}_{j\neq k}\rangle^{{\mbox{\footnotesize co}}}\eqndot
 \end{align}
Let us carry this notation over to operators and denote the $N$th descendant of some operator $\cO$ by $S_-^N[\mathcal{O}]$. Explicitly, for a general operator $\cO$ corresponding to a Bethe state $|\{u_i\}\rangle$
\begin{align}
  \mathcal{O} = \tr\prod_{l=1}^L\Big(\langle\uparrow_l \!\! |\otimes X+\langle\downarrow_l \!\! |\otimes Y\Big) |\{u_i\} \rangle \eqncom
\end{align}
the descendant is defined as
\begin{align}
  S_-^A[\mathcal{O}] = \tr\prod_{l=1}^L\Big(\langle\uparrow_l \!\! |\otimes X+\langle\downarrow_l \!\! |\otimes Y\Big)S_-^A |\{u_i\} \rangle \eqndot
\end{align}
For descendant states with $M$ finite and $N-M$ infinite roots, one has the following expression for the norm~\cite{Escobedo:2010xs}:
\begin{align} \label{eq:NormDesc}
^{{\mbox{\footnotesize co}}}\langle\{u_j,\infty^{N-M}\} |\{u_j,\infty^{N-M}\}\rangle^{{\mbox{\footnotesize co}}}
=\frac{(L-2M)!(N-M)!}{(L-M-N)!} \, \,
^{{\mbox{\footnotesize co}}}\langle\{u_j\} |\{u_j\}\rangle^{{\mbox{\footnotesize co}}}\eqndot
 \end{align}
For the overlap, we find a similar relation:
\begin{align}\label{eq:MPSdesc}
 \langle \mbox{MPS} |\{u_j,\infty^{N-M}\}\rangle^{{\mbox{\footnotesize co}}} =
\frac{(N-M)!(\frac{L}{2}-M)!}{(\frac{N-M}{2})!(\frac{L-M-N}{2})!}
 \langle \mbox{MPS} |\{u_j\}\rangle^{{\mbox{\footnotesize co}}}\eqndot
\end{align}
We have checked the above relation for chains up to $L=18$. In particular, one finds that 
\begin{align}
(S^{+})^{N-M}|\mbox{MPS}\rangle = \frac{(N-M)!(\frac{L}{2}-M)!}{(\frac{N-M}{2})!(\frac{L-M-N}{2})!} |\mbox{MPS}\rangle  + S^{-}|\cdots\rangle\eqndot
\end{align}
where the second term vanishes upon taking the inner product with a Bethe state since Bethe states are highest weight states.

Summarizing, from \eqref{eq:NormDesc} and \eqref{eq:MPSdesc} we see that the one-point functions of descendant operators are proportional to the original operators. The proportionality factor is a simple combinatorical factor depending on $L,M,N$.

\section{Rewriting one-point functions\label{App:Rewriting}}
\label{app: one-point functions}

In this appendix, we will show that the tree-level one-point functions of operators from the SU(2) sector are polynomial in $k$. From the closed formula given in our earlier work \cite{Buhl-Mortensen:2015gfd}, this nature of the $k$ dependence is not apparent. More precisely, the one-point function is given by
\begin{align}\label{eq:OnePoint}
C_k = 2^{L-1}C_2 \sum_{j=\frac{1-k}{2}}^{\frac{k-1}{2}} j^L \prod_{i=1}^{M/2} \frac{u_i^2(u_i^2+\frac{k^2}{4})}{[u_i^2 + (j-\half)^2][u_i^2 + (j+\half)^2]}\eqncom
\end{align}
where $C_2$ is the one-point function for $k=2$. In general, $C_k$ will depend rationally on $k$, but we will show that the dependence becomes polynomial on solutions of the Bethe equations. The reason that this happens is that the above proportionality factor is given by the SU$(2)$ transfer matrix in the $k$-dimensional representation \cite{ZaremboTalk}.
Let us briefly review the form of the transfer matrix of the SU$(2)$ spin chain using Baxter polynomials following \cite{Fioravanti:2016bmi}.

\paragraph{Transfer matrix and Baxter polynomials} Define the Baxter polynomial of degree $M$ as
\begin{align}
Q(u) = \prod_{i=1}^M (u-u_i)\eqndot
\end{align}
The transfer matrix in the fundamental representation $T_1$ satisfies the so-called Baxter TQ-relation
\begin{align}
T_1(u) Q(u) = (u-\tfrac{i}{2})^L Q(u+i)+  (u+\tfrac{i}{2})^L Q(u-i)\eqndot
\end{align}
Since $Q(u_j)=0$ by construction, analyticity of $T_1$ implies that 
\begin{align}\label{eq:BAE}
&0=(u_j-\tfrac{i}{2})^L Q(u_j+i)+  (u_j+\tfrac{i}{2})^L Q(u_j-i)
&&\Rightarrow &&\left[\frac{u_j+\frac{i}{2}}{u_j-\frac{i}{2}}\right]^L = \prod_{i\neq j}\frac{u_j-u_i+i}{u_j- u_i-i}\eqncom
\end{align}
which are the Bethe equations. Assuming $Q$ to be real analytic, we then can recursively relate the transfer matrix for the $(n+1)$-dimensional representation 
to the one of the $n$-dimensional one as follows:
\begin{align}
T_n(u) Q\Big(u+\tfrac{i(n-1)}{2}\Big) - T_{n-1} \Big(u-\tfrac{i}{2} \Big) Q \Big(u+\tfrac{i(n+1)}{2} \Big) = \Big(u+\tfrac{in}{2} \Big)^L Q \Big(u-\tfrac{i(n+1)}{2} \Big)\eqndot
\end{align}
This can be recursively solved: 
\begin{align}\label{eq:Transfer}
T_n(u) =  \sum_{a=-\frac{n}{2}}^{\frac{n}{2}} (u+ia)^L\frac{Q(u+\frac{n+1}{2}i)Q(u-\frac{n+1}{2}i)}{Q(u+(a-\frac{1}{2})i)Q(u+(a+\frac{1}{2})i)}\eqndot
\end{align}
Coming back to our formula of one-point functions, recall that all rapidities are paired, i.e.\ $u_i=-u_{M/2+i}$. This implies that we can write \eqref{eq:OnePoint} in terms of Baxter polynomials 
\begin{align}\label{eq:OnePointQ}
C_k = 2^{L-1}C_2 \sum_{j=\frac{1-k}{2}}^{\frac{k-1}{2}} j^L \frac{Q(0)Q(\frac{ik}{2})}{Q( (j-\frac{1}{2})i)Q((j+\frac{1}{2})i)}\eqndot
\end{align}
Comparing this against \eqref{eq:Transfer}, we immediately see that
\begin{align}
C_k = (2i)^{L} \frac{Q(0)}{Q(\frac{ik}{2})}\frac{T_{k-1}(0)}{2} C_2\eqndot
\end{align}

\paragraph{Baxter polynomials} Next, we make the $k$ dependence of the one-point function explicit. To that end, we notice that the product of Baxter polynomials in the denominator of the transfer matrix can be partially fractioned via
\begin{align}
 \frac{j^L}{Q((j-\half)i)Q((j+\half)i)} = 
-\sum_{i=1}^{M/2} \frac{1}{Q^\prime(u_i)} \Biggl( &\frac{u_i+\frac{i}{2}}{Q(u_i+i)} \! \left[\frac{j^{L-1}}{j-i(u_i+\frac{i}{2})}+\frac{j^{L-1}}{j+i(u_i+\frac{i}{2})}\right] 
\\&  -
  \frac{u-\frac{i}{2}}{Q(u_i-i)}  \left[\frac{j^{L-1}}{j-i(u_i-\frac{i}{2})}+\frac{j^{L-1}}{j+i(u_i-\frac{i}{2})}\right] \Biggr) \nonumber\eqncom
\end{align}
where we again used the fact that the rapidities are paired.
Each term can be further simplified using the identity
\begin{align}
\sum_{j=\frac{1-k}{2}}^{\frac{k-1}{2}} \frac{j^{L-1}}{j-a} = a^{L-1}\left[\Psi(\tfrac{1-k}{2}-a)-\Psi(\tfrac{1+k}{2}-a) \right]-2 \sum_{m=1}^{L/2} a^{L-2m}\mathcal{B}_{2m-1}\eqndot
\end{align}
Using the fact that $L$ is even, we also have that
$
\sum \frac{j^{L-1}}{j-a} = \sum \frac{j^{L-1}}{j+a}\eqndot
$
This implies
\begin{align}
\sum_{j=\frac{1-k}{2}}^{\frac{k-1}{2}} \frac{j^L}{Q((j-\half)i)Q((j+\half)i)} &= 
\sum_i\frac{4i^L}{Q^\prime(u_i)}  \bigg\{
\frac{i}{2}\frac{(u_i+\frac{i}{2})^{L}}{Q(u_i+i)}\frac{k}{u_i^2+ \frac{k^2}{4}}  \\\nonumber
&+\sum_{m=1}^{L/2} \left[\frac{(u_i+\frac{i}{2})^{L-2m+1}}{Q(u_i+i)}+\frac{(u_i-\frac{i}{2})^{L-2m+1}}{Q(u_i-i)} \right] 
\frac{\mathcal{B}_{2m-1}}{i^{2m}} \\
&- \frac{i}{2}\left[\frac{(u_i+\frac{i}{2})^{L}}{Q(u_i+i)}+\frac{(u_i-\frac{i}{2})^{L}}{Q(u_i-i)} \right] \!\!\! \left[\Psi(-\tfrac{k}{2}-iu_i )-\Psi(\tfrac{k}{2}-iu_i) \right]\bigg\}\eqndot\nonumber
\end{align}
Now let us compare the left- and right-hand side of the above equation. In particular, we see that in the limit $u_1\rightarrow\infty$ the left-hand side scales like $u_1^{-4}$. In order for the above equation to hold, this means that the right-hand side must display the same behaviour. It is easy to see that this implies that the sum in the second line only runs up to $\frac{L}{2}-M+1$. One indeed quickly checks that the coefficients in front of the Bernoulli polynomials with higher indices vanish. Then, upon using the Bethe equations \eqref{eq:BAE}, we arrive at
\begin{multline}
\sum_{j=\frac{1-k}{2}}^{\frac{k-1}{2}} \frac{j^L}{Q((j-\half)i)Q((j+\half)i)}  =\\ 
\sum_i\frac{4i^L}{Q^\prime(u_i)}  \frac{(u_i+\frac{i}{2})^{L}}{Q(u_i+i)}\Biggl\{ \frac{i}{2}\frac{k}{u_i^2+ \frac{k^2}{4}}   
+\sum_{m=1}^{\frac{L}{2}-M+1} \left[\frac{1}{(u_i+\frac{i}{2})^{2m-1}}- \frac{1}{(u_i-\frac{i}{2})^{2m-1}} \right] 
\frac{\mathcal{B}_{2m-1}}{i^{2m}}\Biggr\}\eqndot 
\end{multline}
Define the conserved charges $q_r$ in the standard way as
\begin{align}
q_r = \frac{i}{r-1} \left[ \frac{1}{(u+\frac{i}{2})^{r-1}} -\frac{1}{(u-\frac{i}{2})^{r-1}}\right]\eqndot
\end{align}
Then,
\begin{align}
&\sum_{j=\frac{1-k}{2}}^{\frac{k-1}{2}} \frac{j^L}{Q((j-\half)i)Q((j+\half)i)} = \,\nonumber\\
&\qquad \qquad \sum_i\frac{4i^L}{Q^\prime(u_i)}  \frac{(u_i+\frac{i}{2})^{L}}{Q(u_i+i)}\left[ \frac{i}{2}\frac{k}{u_i^2+ \frac{k^2}{4}}  -
\sum_{m=1}^{\frac{L}{2}-M+1} \frac{(2m-1)\, q_{2m}(u_i)}{i^{2m-1}}  \mathcal{B}_{2m-1}\right]\eqndot
\end{align}
We can now insert this into \eqref{eq:OnePointQ} to obtain
\begin{align}
C_k=2 C_2
(2i)^L\sum_i\frac{Q(0)}{Q^\prime(u_i)}  \frac{(u_i+\frac{i}{2})^{L}}{Q(u_i+i)}\Biggl[ \frac{\frac{ik}{2}\, Q(\frac{ik}{2})}{u_i^2+ \frac{k^2}{4}}  -
Q(\tfrac{ik}{2})\sum_{m=1}^{\frac{L}{2}-M+1} \frac{(2m-1)\, q_{2m}(u_i)}{i^{2m-1}}  \mathcal{B}_{2m-1}\Biggr].
\end{align}
Notice that both terms are polynomial in $k$ since $Q(\frac{ik}{2}) = \prod_i [u_i^2 +\frac{k^2}{4} ]$. We see that the one-point function is a polynomial of degree $L-M+1$.

\section{Fuzzy spherical harmonics and their products\label{App:Fuzzy}}
In this appendix, our conventions for the fuzzy spherical harmonics $\hat{Y}_\ell^m$ with $\ell=0,\dots,k-1$ and $m=-\ell,\dots,+\ell$ are laid out. 
In evaluating traces, we exploit a number of useful identities for the fuzzy spherical harmonics including formulas for the expansion coefficients of the SU$(2)$ generators $t_i$ in terms of $\hat{Y}_\ell^m$.

The fuzzy spherical harmonics $\hat{Y}^m_\ell$ of dimension $k\times k$ are given by $\hat{Y}_\ell^m = [ \hat{Y}^m_\ell]_{n,n'}  E^{n}_{n'}$, where the matrix elements are \cite{Hoppe82,Kawamoto:2015qla}
\begin{equation}
[ \hat{Y}^m_\ell]_{n,n'} = (-1)^{k-n} \sqrt{2\ell+1} \begin{pmatrix} \frac{k-1}{2} & \ell & \frac{k-1}{2} \\ n-\frac{k+1}{2} & m & -n'+\frac{k+1}{2} \end{pmatrix} \, , \quad n,n' = 1,\dots,k
\end{equation}
and the parenthesis denotes a Wigner 3$j$ symbol.
They are normalised to satisfy $(\hat{Y}^m_\ell)^\dagger = (-1)^m \hat{Y}^{-m}_\ell$ and $\tr (\hat{Y}^{m_1}_{\ell_1} (\hat{Y}^{m_2}_{\ell_2})^\dagger) = \delta_{\ell_1, \ell_2} \delta_{m_1,m_2}$.
The product of fuzzy spherical harmonics can again be expanded in fuzzy spherical harmonics:
\begin{equation}
\hat{Y}_{\ell_1}^{m_1}\hat{Y}_{\ell_2}^{m_2} = \sum_{\ell_3=0}^{k-1} \sum_{m_3=-\ell_3}^{\ell_3} F_{\ell_1 m_1 \ell_2 m_2}^{\ell_3 m_3} \hat{Y}_{\ell_3}^{m_3} \,,
\end{equation}
with fusion coefficients
\begin{equation}
\label{eq:fusioncoefficient}
\begin{aligned}
 F_{\ell_1 m_1 \ell_2 m_2}^{\ell_3 m_3} &= (-1)^{\ell_1 +\ell_2+\ell_3+m_3}\sqrt{(2\ell_1+1)(2\ell_2+1)(2\ell_3+1)} \\
 &\phaneq\times\begin{pmatrix} \ell_1 & \ell_2 & \ell_3 \\ m_1 & m_2 & -m_3\end{pmatrix} \begin{Bmatrix} \ell_1 & \ell_2 & \ell_3 \\ \frac{k-1}{2} & \frac{k-1}{2} & \frac{k-1}{2}  \end{Bmatrix}\, ,
\end{aligned}
\end{equation}
where the curly bracket denotes a 6$j$ symbol. 
The fuzzy spherical harmonics thus satisfy an algebra, the fusion algebra of fuzzy spherical harmonics \cite{Kawamoto:2015qla}.
Since the fusion coefficients $F_{\ell_1 m_1 \ell_2 m_2}^{\ell_3 m_3}$ have a Wigner 3$j$ symbol as a factor, it is useful to be recall the selection rules for Wigner 3$j$ symbols, which are essentially addition of angular momenta. Consider the 3$j$ symbol in the fusion coefficients
\begin{equation}
 \begin{pmatrix} \ell_1 & \ell_2 & \ell_3 \\ m_1 & m_2 & -m_3\end{pmatrix} \eqndot
\end{equation}
This is zero unless: (1) $m_i$ is one of the values $-\ell_i, -\ell_i+1, \dots, \ell_i-1, \ell_i$, (2) the $\ell_i$'s satisfy the triangular condition $\abs{\ell_1-\ell_2} \leq \ell_3 \leq \ell_1+\ell_2$ and (3) $m_3 = m_1 + m_2$. Besides, (4) $\ell_1+\ell_2+\ell_3$ must be an integer and further an even integer if all the magnetic quantum numbers are zero $m_1=m_2=m_3 = 0$.%

\paragraph{Trace formulas and $\alpha_m^L$ coefficients} 
Since the fuzzy spherical harmonics span the space of $k\times k$ matrices, any matrix can be decomposed as an expansion in $\hat{Y}_\ell^m$ \cite{Kawamoto:2015qla}. 
In our conventions, the generators have the expansion
\begin{align} 
t_1 &= \frac{(-1)^{k+1}}{2} \sqrt{ \frac{k(k^2-1)}{6}} (\hat{Y}_1^{-1}-\hat{Y}_1^1) \eqncom\\
t_2 &= i \frac{(-1)^{k+1}}{2} \sqrt{ \frac{k(k^2-1)}{6}} (\hat{Y}_1^{-1}+\hat{Y}_1^1)\eqncom\\
t_3 &= \frac{(-1)^{k+1}}{2} \sqrt{ \frac{k(k^2-1)}{3}} \hat{Y}_1^0 \eqndot
\end{align}
Any trace of products of $t_i$s and $\hat{Y}_\ell^m$s can then be expressed in terms of sums of products of the fusion coefficients defined above \eqref{eq:fusioncoefficient}. 

The expansion of $t_3^L$ can be found from the expansion of $t_3 \sim \hat{Y}_1^0$ using the fusion algebra of fuzzy spherical harmonics. The expansion takes the form
\begin{equation}
\label{eq: definition of alpha}
t_3^L = \sum_{\ell=0}^{L} \alpha_\ell^L \hat{Y}_\ell^0 \, ,
\end{equation}
where the coefficients $\alpha_\ell^L$ are then given by
\begin{equation}
\alpha_\ell^L = \left( \frac{(-1)^{k+1}}{2} \sqrt{\frac{k(k^2-1)}{3}} \right)^L \sum_{\lambda_1}\dots\sum_{\lambda_{L-2}}  F_{1010}^{\lambda_1 0} \prod_{i=1}^{L-3} F_{\lambda_{i}0 1 0}^{\lambda_{i+1}0} F_{\lambda_{L-2}010}^{\ell 0} \eqndot \label{eq:appfuzzy:alpha}
\end{equation}
The possible form follows from the selection rules of 3$j$ symbols from which we further see that only even (odd) $\ell$ contributes for even (odd) $L$.

In general, the formula \eqref{eq:appfuzzy:alpha} for $\alpha_\ell^L$ contains many terms: one term for each lattice walk from 1 to a given number $0 \leq \ell \leq L$ in $L-2$ steps.
In the following, we reduce the computational complexity from exponential in $L$ to polynomial in $L$; the resulting expressions are given in \eqref{eq: result alphas}.

If we are interested in $\alpha^L_L$, i.e.\ the coefficient of the highest spin fuzzy harmonic in the expansion of $t_3^L$, the sum reduces to a single term:
\begin{equation}
\alpha_L^L = \frac{(-1)^{(k+1)L}}{2^L}\left(\frac{k(k^2-1)}{3}\right)^\frac{L}{2} F_{1010}^{20} F_{2010}^{30} \dots F_{L-1010}^{L0}  \eqndot
\end{equation}
By inserting the explicit expressions for the fusion coefficients and the Wigner symbols therein, one finds a simple formula for the coefficient of the highest spin contribution to $t_3^L$:
\begin{equation}
\alpha_L^L = \frac{(-1)^{k+1} }{\sqrt{2L+1}\binom{2L}{L}}  \sqrt{\frac{\Gamma(k+L+1)}{\Gamma(k-L)}}\eqndot
\label{eq:appfuzzy:alphaLL}
\end{equation}

For $L$ even, $\tr(t_3^L) = -2\mathcal{B}_{L+1}$ with $\mathcal{B}$ defined in \eqref{eq: definition curly B}. Furthermore, the trace of a fuzzy spherical harmonic is zero for all $\ell, m$ except $\ell=m = 0$, which has the trace $\tr(\hat{Y}_0^0)=(-1)^{k+1} \sqrt{k}$.
Therefore, the coefficient of the lowest spin fuzzy harmonic $\ell=m=0$ in the expansion of $t_3^L$ is easy to evaluate
\begin{equation}
-2\mathcal{B}_{L+1} = \tr(t_3^L) = \sum_{\ell=0}^{L} \alpha_\ell^L \tr(\hat{Y}_\ell^0) = \alpha_0^L (-1)^{k+1} \sqrt{k}\eqndot
\end{equation}
Thus,
\begin{equation}
\alpha_0^L = 2 \frac{(-1)^k}{\sqrt{k}}\mathcal{B}_{L+1} \eqndot
\end{equation}

This fact can be exploited to produce a numerically efficient recursion relation for the coefficients. Let $L+m$ be even but $m$, $L$ otherwise arbitrary. Since the fuzzy spherical harmonics are orthogonal, we have
\begin{align}
-2\mathcal{B}_{L+m+1} = \tr(t_3^{L+m}) = \tr(t_3^{L}t_3^{m}) = \sum_{\ell=0}^{\min(L,m)} \alpha_\ell^L \alpha_\ell^m\eqndot
\end{align}
Assume without loss of generality that $m \leq L$. Thus,
\begin{equation}
\sum_{\ell=0}^{m} \alpha_\ell^L \alpha_\ell^m = \alpha^L_m \alpha^m_m + \sum_{\ell=0}^{m-1} \alpha_\ell^L \alpha_\ell^m  = -2\mathcal{B}_{L+m+1} 
\end{equation}
or in other words
\begin{equation}
\alpha_m^L = \frac{1}{\alpha^m_m} \left( -2\mathcal{B}_{L+m+1} - \sum_{\ell=0}^{m-1} \alpha_\ell^L \alpha_\ell^m \right)
\eqndot
\label{eq:alpharecursion}
\end{equation}
Note that if $L+m$ was odd we would have gotten zero for the trace. This is a recursion relation for the coefficients as it only depends on $\alpha^\ell_p$ for $\ell<L$ and $p<m$ apart from the $\alpha^m_m$ which we know from the formula \eqref{eq:appfuzzy:alphaLL} above. 

The recursion can be solved by an ansatz for the coefficients. Note that all the $L$ dependence comes from the $\alpha^L_\ell$ factor and that by reinserting the recursion this comes in the form of $\mathcal{B}_{L+\ell+1}$. 
Thus, we obtain the form 
\begin{equation}
\alpha_m^L = \frac{2}{\alpha_m^m} \Bigl( -\mathcal{B}_{L+m+1} + \sum_{\ell=0}^{m-1} \beta^{(\ell)}_m \mathcal{B}_{L+\ell+1} \Bigr),
\end{equation}
with coefficients $\beta^{(\ell)}_m = 0$ for $\ell$ even (odd) if $m$ is odd (even). Then, for a given $m$, there is one $\beta^{(\ell)}_m$ for each odd (even) number less than $m$. Furthermore, $\alpha_m^L = 0$ for $m>L$, which in fact gives an equation for each odd (even) $L$ less than $m$. 
Therefore, the $\beta$s are in fact fixed by this requirement.

Specialising to odd $L$ and $m$ gives the equations
\begin{equation}
\mathcal{B}_{L+m+1} = \beta^{(1)}_m \mathcal{B}_{L+2} + \beta^{(3)}_m \mathcal{B}_{L+4} + \dots + \beta^{(m-2)}_m \mathcal{B}_{L+m-1}
\end{equation}
for $L=1,3,\dots,m-2$. These are linear equations and can be represented as a matrix equation for a vector $\beta_m = (\beta^{(1)}_m, \beta^{(3)}_m, \dots, \beta^{(m-2)}_m)$:
\begin{equation}
\begin{pmatrix}
\mathcal{B}_3 & \mathcal{B}_5 & \dots & \mathcal{B}_{m} \\
\mathcal{B}_5 & \mathcal{B}_7 & \dots & \mathcal{B}_{m+2} \\
\vdots & & \ddots & \\
\mathcal{B}_{m} & \mathcal{B}_{m+2} & \dots & \mathcal{B}_{2m-4}
\end{pmatrix} 
\beta_m
=
\begin{pmatrix}
\mathcal{B}_{m+2} \\
\mathcal{B}_{m+4} \\
\vdots \\
\mathcal{B}_{2m-1}
\end{pmatrix}
\eqndot
\end{equation}

For even $L$ and $m$, the equations are
\begin{align}
\mathcal{B}_{L+m+1} =  \beta^{(0)}_m \mathcal{B}_{L+1} + \beta^{(2)}_m \mathcal{B}_{L+3} + \dots + \beta^{(m-2)}_m \mathcal{B}_{L+m-1}, 
\end{align}
for $L=0,2,\dots,m-2$. These are also linear equations and can be represented as a matrix equation for a vector $\beta_m = (\beta^{(0)}_m, \beta^{(2)}_m, \dots, \beta^{(m-2)}_m)$:
\begin{equation}
\begin{pmatrix}
\mathcal{B}_1 & \mathcal{B}_3 & \dots & \mathcal{B}_{m-1} \\
\mathcal{B}_3 & \mathcal{B}_5 & \dots & \mathcal{B}_{m+1} \\
\vdots & & \ddots & \\
\mathcal{B}_{m-1} & \mathcal{B}_{m+1} & \dots & \mathcal{B}_{2m-3}
\end{pmatrix} 
\beta_m
=
\begin{pmatrix}
\mathcal{B}_{m+1} \\
\mathcal{B}_{m+3} \\
\vdots \\
\mathcal{B}_{2m-1}
\end{pmatrix}
\eqndot
\end{equation}

Thus defining the coefficient matrix as $M_m$ and the right-hand side as $b_m$, we can define another vector $\tilde{\beta}_m = (M_m^{-1} b_m, -1)$ in terms of which the $\alpha_m^L$ introduced in \eqref{eq: definition of alpha} can be written as
\begin{equation}
\label{eq: result alphas}
\alpha_m^L =2 (-1)^{k+1} \sqrt{2m+1}\binom{2m}{m} \sqrt{\frac{\Gamma(k-m)}{\Gamma(k+m+1)}} \sum_{\ell=0}^{m} \tilde{\beta}^{(\ell)}_m \mathcal{B}_{L+\ell+1} \, ,
\end{equation}
where the sum is over all even (odd) numbers $0\leq \ell \leq m$ for even (odd) $L$.

Finally, let us give the large $k$ expansion
\begin{align}\label{eq:alphaK}
\alpha^L_m =
(-1)^k \left(\frac{k}{2}\right)^{L+\half}\sqrt{m+\half}i^m
\left\{ 
\begin{array}{ll}
\frac{L}{L+1} \frac{\big(\frac{2-L}{2}\big)_{\frac{m-2}{2}}}{\big(\frac{L+3}{2}\big)_{\frac{m}{2}}}& L,m~ \mathrm{even,}\\
i \frac{\big(\frac{1-L}{2}\big)_{\frac{m-1}{2}}}{\big(\frac{L+2}{2}\big)_{\frac{m+1}{2}}} & L,m~ \mathrm{odd,}
\end{array}\right.
\end{align}
where $(a)_n$ is the Pochhammer symbol.

The other generators have a more involved expansion; however, in general all the coefficients are related to the $\alpha$s as
\begin{align}
&\tr(t_1^L  \hat{Y}_\ell^m) =(-1)^\frac{m-2L}{2} \tr(t_2^L  \hat{Y}_\ell^m)=\frac{(-1)^\frac{\ell+m}{2}}{2^{\ell}} \sqrt{\frac{(\ell+m)!}{(\ell-m)!}} \frac{(\ell-m)!}{(\frac{\ell-m}{2})!(\frac{\ell+m}{2})!} \tr(t_3^L  \hat{Y}_l^0) 
\end{align}
for $L, \ell, m$ all even or all odd; otherwise the traces gives zero.

\section{Boundary operators}
\label{app: boundary operators}

In this appendix, we derive the spectrum of boundary operators that can occur at the defect.
We first need to understand what fields we have available. On the defect, there is
a dynamical 3d hypermultiplet consisting of the scalar $\fscal$ and the fermion $\fferm$~\cite{DeWolfe:2001pq}.
Both are in the fundamental of the $\text{U}(N-k)$ 
gauge group. An additional
class of boundary fields is obtained by taking suitable limits of the $x_3 > 0$ bulk
fields, as we will now explain.

\subsection{Gauge-covariant boundary fields}
Due to the $x_3$-dependent mass terms, the fields outside the $(N-k)^2$ block fall
of with some power of $x_3$ near the defect. From the explicit propagator \eqref{eq:Knu}
and the masses in table \ref{tab:spectrum}, we find
\begin{equation}
  (\scalq_{4,5,6},A_{0,1,2},c)_{\ell m}(x) \sim (x_3)^{\ell+1}\eqncom\qquad
  (\scalq_{1,2,3},A_3,\ferm_{1,2,3,4})_{\ell m}(x) \sim (x_3)^\ell\eqncom
\end{equation}
and ($n = 1,\ldots,k$ and $a = k+1,\ldots,N$ are colour indices)
\begin{equation}
  [\scalq_{4,5,6},A_{0,1,2},c]_{n,a}(x) \sim (x_3)^{\frac{k+1}{2}}\eqncom\qquad
  [\scalq_{1,2,3},A_3,\ferm_{1,2,3,4}]_{n,a}(x) \sim (x_3)^{\frac{k-1}{2}}\eqncom
\end{equation}
as $x_3 \to 0^+$. 
We can thus define finite fields on the defect (denoted by adding a hat) as 
a limit scaled by the appropriate power of $x_3$, e.g.\
\begin{equation}
  ({\hat\phi}_1)_{\ell,m}(\vec{x}) = \lim_{x_3 \to 0^+}(x_3)^{-\ell}(\scalq_1)_{\ell,m}(x) \eqndot
\end{equation}

In order to construct physical operators, it is useful to have a basis of fields
that transform in a simple way under the gauge symmetry.
The BRST-variation of the bulk fields are \cite{Buhl-Mortensen:2016jqo}
\begin{equation}
  sA_\mu = \cder_\mu c = \partial_\mu c -i[A_\mu,c]\eqncom \quad
    s\scalq_i = -i[\phi_i,c]\eqncom \quad
    s\psi_i = i\{\psi_i,c\}\eqncom
\end{equation}
and the variation of the boundary fields follows by taking the $x_3 \to 0$ limit.
Let us consider $(\hat{\phi}_{1,2,3})_{\ell m}$ as an example. We have
\begin{equation}
\begin{aligned}
  s(\scalq_i)_{\ell m} &= 
    \frac{i}{x_3}c_{\ell'm'}\tr\left([t_i,\hat{Y}_{\ell'}^{m'}](\hat{Y}_\ell^m)^\dagger\right) 
    -i(\scalq_i)_{\ell_1 m_1}c_{\ell_2 m_2}
        \tr\left([\hat{Y}_{\ell_1}^{m_1},\hat{Y}_{\ell_2}^{m_2}](\hat{Y}_\ell^m)^\dagger\right)\\
        &\phaneq -i([\scalq_i]_{n,a}[c]_{a,n'}-[c]_{n,a}[\scalq_i]_{a,n'})
                                [(\hat{Y}_\ell^m)^\dagger]_{n',n}\eqndot
\end{aligned}
\end{equation}
The first term can be simplified using that
\begin{equation}
  \tr\left([t_i,\hat{Y}_{\ell'}^{m'}](\hat{Y}_\ell^m)^\dagger\right)
    = \delta_{\ell \ell'}[t_i^{(2\ell+1)}]_{\ell-m+1,\ell-m'+1}\eqncom
\end{equation}
and is then seen to have a finite limit. The second term vanishes in the $x_3\to 0^+$ limit because the fusion rules
(see appendix \ref{App:Fuzzy}) imply that $\ell_1+\ell_2 \geq \ell$; likewise, the third term does not contribute 
because $k > \ell$.
The result is shown in table \ref{tab:bdry BRST}, along with the variation of the remaining
boundary fields.

\begin{table}
\begin{center}
\begin{tabular}{c|c}
  $\Phi$                             & $s\Phi$ \\ \hline
  $({\hat\phi}_{1,2,3})_{\ell m}$    & $i[t^{(2\ell+1)}_{1,2,3}]_{\ell-m+1,\ell-m'+1}{\hat c}_{\ell m'}$ \\
  $({\hat\phi}_{4,5,6})_{\ell m}$    & 0 \\
  $({\hat A}_{\hat\mu})_{\ell m}$    & $\partial_{\hat\mu}{\hat c}_{\ell m}$ \\
  $({\hat A}_3)_{\ell m}$            & ${\hat c}_{\ell m}$ \\
  $({\hat\ferm}_{1,2,3,4})_{\ell m}$ & 0 \\ \hline
  $[{\hat\phi}_{1,2,3}]_{n,a}$       & $-i[{\hat\phi}_{1,2,3}]_{n,a'}[c]_{a',a} + i[t_i]_{n,n'}[\hat{c}]_{n',a}$ \\
  $[{\hat\phi}_{4,5,6}]_{n,a}$       & $-i[{\hat\phi}_{4,5,6}]_{n,a'}[c]_{a',a} + i[\hat{c}]_{n,a'}[\phi_{4,5,6}]_{a',a}$ \\
  $[{\hat A}_{\hat\mu}]_{n,a}$       & $\partial_{\hat\mu}[{\hat c}]_{n,a} - i[\hat{A}_{\hat\mu}]_{n,a'}[c]_{a',a} + i[\hat{c}]_{n,a'}[A_{\hat\mu}]_{a',a}$ \\
  $[{\hat A}_3]_{n,a}$               & $[\hat c]_{n,a} - i[\hat{A}_3]_{n,a'}[c]_{a',a}$ \\
  $[{\hat\ferm}_{1,2,3,4}]_{n,a}$    & $i[\hat{\ferm}_{1,2,3,4}]_{n,a'}[c]_{a',a}$
\end{tabular}
\end{center}
\caption{BRST variation of boundary fields.}
\label{tab:bdry BRST}
\end{table}

The terms in the gauge variation involving $\hat{c}_{\ell m}$ and $[\hat{c}]_{n,a}$
can be eliminated by using $\hat{A}_3$ to construct covariant fields. Explicitly, we 
redefine
\begin{equation}
  (\hat{\phi}_{1,2,3})_{\ell m} \to (\hat{\phi}_{1,2,3})_{\ell m}
    -i[t^{(2\ell+1)}_{1,2,3}]_{\ell-m+1,\ell-m'+1} (\hat{A}_3)_{\ell m'}\eqncom \quad
  (\hat{A}_{\hat \mu})_{\ell m} \to (\hat{A}_{\hat \mu})_{\ell m} 
                                      - \partial_{\hat\mu}(\hat{A}_3)_{\ell m}\eqncom
\end{equation}
\begin{equation}
  [\hat{\phi}_{1,2,3}]_{n,a} \to [\hat{\phi}_{1,2,3}]_{n,a} - i[t_i]_{n,n'}[\hat{A}_3]_{n',a}\eqncom\quad
  [\hat{\phi}_{4,5,6}]_{n,a} \to [\hat{\phi}_{4,5,6}]_{n,a} 
                                - i[\hat{A}_3]_{n,a'}[\phi_{4,5,6}]_{a',a}\eqncom
\end{equation}
and
\begin{equation}
  [\hat{A}_{\hat\mu}]_{n,a} \to [\hat{A}_{\hat\mu}]_{n,a} - \partial_{\hat\mu}[\hat{A}_3]_{n,a} - i[\hat{A}_3]_{n,a'}[\hat{A}_{\hat\mu}]_{a',a} \eqndot
\end{equation}
We thus obtain the fields listed in table \ref{tab:bdry-fields}, which are
either gauge singlets, or in the fundamental representation of U$(N-k)$, as expected.

\begin{table}
\begin{center}
\begin{tabular}{c|c|c|c|c}
  $\Phi$                             & $\hat{\Delta}$       & $\soc$                             & $\soe$        & U$(N-k)$ \\ \hline
  $({\hat\phi}_{1,2,3})_{\ell m}$   & $\ell + 1$           & $1\otimes \ell$                    & 0             & singlet \\
  $({\hat\phi}_{4,5,6})_{\ell m}$    & $\ell + 2$           & $\ell$                             & 1             & singlet \\
  $({\hat A}_{\hat\mu})_{\ell m}$   & $\ell + 2$           & $\ell$                             & 0             & singlet \\
  $({\hat\ferm}_{1,2,3,4})_{\ell m}$ & $\ell + \frac{3}{2}$ & $\frac{1}{2}\otimes \ell$          & $\frac{1}{2}$ & singlet \\ \hline
  $[{\hat\phi}_{1,2,3}]_{n,a}$      & $\frac{k+1}{2}$      & $1\otimes \frac{k-1}{2}$           & 0             & fundamental \\
  $[{\hat\phi}_{4,5,6}]_{n,a}$      & $\frac{k+3}{2}$      & $\frac{k-1}{2}$                    & 1             & fundamental \\
  $[{\hat A}_{\hat\mu}]_{n,a}$      & $\frac{k+3}{2}$      & $\frac{k-1}{2}$                    & 0             & fundamental \\
  $[{\hat\ferm}_{1,2,3,4}]_{n,a}$    & $\frac{k+2}{2}$      & $\frac{1}{2}\otimes \frac{k-1}{2}$ & $\frac{1}{2}$ & fundamental \\ \hline
  $\fscal_a$                         & $\frac{1}{2}$        & $\frac{1}{2}$                      & 0             & fundamental \\
  $\fferm_a$                         & 1                    & 0                                  & $\frac{1}{2}$ & fundamental
\end{tabular}
\end{center}
\caption{Boundary fields.}
\label{tab:bdry-fields}
\end{table}

The defect breaks the SO$(6)$ $R$-symmetry down to $\soc\times\soe$.
Under $\soe$, the boundary fields transforms in the same way as the bulk fields.
However, the naive action of $\soc$,
\begin{equation}
  \scal_i \to R_{ij} \scal_j\eqncom\qquad i,j=1,2,3\eqncom\quad R \in \text{SO}(3)\eqncom
\end{equation} 
does not preserve our boundary conditions
\begin{equation}
  \scal_i \sim -\frac{t_i}{x_3}\eqncom\qquad \text{as $x_3 \to 0^+$ .}
\end{equation}
This problem can be remedied by defining a `twisted' symmetry by (here $\tilde{R}$ is the
matrix in the $\text{SU}(2)_C$ subgroup of $\text{SU}(4)$ corresponding to $R$)
\begin{equation}
  \scal_i \to U R_{ij} \scal_j U^{-1} \eqncom\qquad i,j=1,2,3\eqncom
  \label{eq:twisted-R1}
\end{equation}
\begin{equation}
  \ferm_i \to U \tilde{R}_{ij} \ferm_j U^{-1}\eqncom
\end{equation}
and 
\begin{equation}
  \Phi \to U \Phi U^{-1}\eqncom \qquad \Phi\in\{\scal_{4,5,6},A_{0,1,2,3},c\}\eqncom
  \label{eq:twisted-R2}
\end{equation}
where $U = \e^{i\alpha_i t_i}$
such that the combined 
transformation preserves the boundary conditions, i.e.
\begin{equation}
  U R_{ij} t_j U^{-1} = t_i \eqndot
\end{equation}
In the bulk, $U$ acts as a constant gauge transformation and is thus irrelevant. In contrast, 
the gauge group is reduced to U$(N-k)$ on the boundary, and the action of $U$ becomes important.
As a result of this twisting, boundary fields fall in a tensor-product representation under
$\soc$, with one factor from the flavour index, and one factor from colour index; see
table \ref{tab:bdry-fields}.

\subsection{Low-dimensional operators}
We are now ready to construct gauge-invariant operators using our boundary fields.
Adapting the language from the bulk theory, we call a boundary operator `multi-trace' if
it is the product of several operators,
\begin{equation}
  \hat{\mathcal O}(\vec x) = \hat{\mathcal O}_1(\vec x)\cdots\hat{\mathcal O}_2(\vec x)\eqncom\qquad n > 1\eqncom
\end{equation}
where each factor $\hat{\mathcal O}_j(\vec x)$ is separately gauge invariant. Otherwise, it
is called `single-trace'. The spectrum of scalar single-trace operators with dimension 
$\hat{\Delta} \leq 2$ is shown in table \ref{tab:single trace boundary}.
By combining two dimension one operators we obtain additional dimension two operators.
The number of double trace multiplets is listed in table \ref{tab:double trace boundary}.
By the same combinatorics as in the bulk case, on can check that mixing between
single- and multi-trace operators is suppressed in the planar limit.

\begin{table}
\begin{center}
\begin{tabular}{c|c|c}
   & $\hat{\Delta} = 1$                                                      & $\hat{\Delta} = 2$ \\ \hline
  $(0,0)$          & $\afscal\fscal$                                                         & $(\hat{\phi}_{1,2,3})_{\ell = 1}$, $\tr[(\phi_{1,2,3})^2]$, $\tr[(\phi_{4,5,6})^2]$, $\afscal\phi_{1,2,3}\fscal$, $\afferm\fferm$ \\
  $(1,0)$          & $(\hat{\phi}_{1,2,3})_{\ell = 0}$, $\tr[\phi_{1,2,3}]$, $\afscal\fscal$ & $(\hat{\phi}_{1,2,3})_{\ell = 1}$, $\tr[D_3\phi_{1,2,3}]$, $\afscal\phi_{1,2,3}\fscal$ \\
  $(2,0)$          &                                                                         & $(\hat{\phi}_{1,2,3})_{\ell = 1}$, $\tr[(\phi_{1,2,3})^2]$, $\afscal\phi_{1,2,3}\fscal$ \\
  $(0,1)$          & $\tr[\phi_{4,5,6}]$                                                     & $(\hat{\phi}_{4,5,6})_{\ell = 0}$, $\tr[D_3\phi_{4,5,6}]$, $\afferm\fferm$, $\afscal\phi_{4,5,6}\fscal$\\
  $(0,2)$          &                                                                         & $\tr[(\phi_{4,5,6})^2]$\\
  $(1,1)$          &                                                                         & $\tr[\phi_{1,2,3}\phi_{4,5,6}]$, $\afscal\phi_{4,5,6}\fscal$
\end{tabular}
\end{center}
\caption{Scalar single-trace operators with $\hat{\Delta} = 1,2$. For brevity we leave out
the explicit group-theoretic coefficients necessary to project out the various irreducible representations,
and assume $k > 2$.
Note that $(1,0)$ occurs twice in the decomposition of $\afscal\phi_{1,2,3}\fscal$.}
\label{tab:single trace boundary}
\end{table}

\begin{table}
\begin{center}
\begin{tabular}{c|c|c|c|c|c|c}
  Multiplet & $(0,0)$ & $(1,0)$ & $(2,0)$ & $(0,1)$ & $(0,2)$ & $(1,1)$ \\ \hline
  \#        & 8       & 6       & 6       & 1       & 1       & 3
\end{tabular}
\end{center}
\caption{Number of scalar double-trace multiplets with $\hat{\Delta} = 2$.}
\label{tab:double trace boundary}
\end{table}

For the calculations in section \ref{sec:BOE}, we need to know which boundary operators the bulk
operator $\tr Z^2$ can couple to. The $\soc\times\soe$ decomposition is 
\begin{equation}
  \tr Z^2 \sim (0,0) \oplus (0,0) \oplus (2,0) \oplus (0,2) \oplus (1,1) \eqndot
\end{equation}
Using the explicit expression given in table \ref{tab:single trace boundary}, we see that, when restricting
to $\hat{\Delta} \leq 2$,
$\tr Z^2$ can only couple to boundary operators in the $[\hat{\Delta}=2,(0,0)]$ and $[\hat{\Delta}=2,(2,0)]$, to leading
order in $g_\YM$.
Neglecting space-time dependence, we thus have
\begin{equation}
  \tr Z^2 \sim \sum_{\alpha = 1}^{13} {\mu_{\tr Z^2}}^{\hat{\mathcal{O}}_{[2,(0,0)],\alpha}}\hat{\mathcal{O}}_{[2,(0,0)],\alpha}
     + \sum_{\beta = 1}^{9} {\mu_{\tr Z^2}}^{\hat{\mathcal{O}}_{[2,(2,0)],0,\beta}}\hat{\mathcal{O}}_{[2,(2,0)],0,\beta}
     + \cdots\eqncom
  \label{eq:trZ2 BOE}
\end{equation}
where the dots denote operators that have dimension higher than two, or which are subleading in $g_\YM$.
For the $[2,(2,0)]$ multiplets, we need to specify which component  appears
in the expansion. Choose a Cartan generator for $\soc$ such that $\scal_3$ has charge zero. Then, 
$\hat{\mathcal{O}}_{[2,(2,0)],r,\beta}$ denotes the component with charge $r$.

Finally, we need to relate ${\mu_{\tr Z^2}}^{\hat{\mathcal{O}}_i}$ and
${\mu_{\tr X^2}}^{\hat{\mathcal{O}}_i}$.
To this end, consider the element of $\soc\times\soe$ which
rotates $\pi/2$ around the 2-axis of $\soc$ and the 5-axis of $\soe$. Acting on \eqref{eq:trZ2 BOE}, we obtain
\begin{equation}
  \tr[X^L] \sim \sum_{\alpha = 1}^{13} {\mu_{\tr Z^2}}^{\hat{\mathcal{O}}_{[2,(0,0)],\alpha}}\hat{\mathcal{O}}_{[2,(0,0)],\alpha}
     + \sum_{\beta = 1}^{9} {\mu_{\tr Z^2}}^{\hat{\mathcal{O}}_{[2,(2,0)],0,\beta}} \sum_{r=-2}^2 c_r \hat{\mathcal{O}}_{[2,(2,0)],r,\beta}
     + \cdots\eqncom
\end{equation}
where $c_r$ is determined by group theory. We can immediately read off that
\begin{equation}
  {\mu_{\tr X^2}}^{\hat{\mathcal{O}}_{[2,(0,0)],\alpha}} = {\mu_{\tr Z^2}}^{\hat{\mathcal{O}}_{[2,(0,0)],\alpha}}\eqncom
\end{equation}
and that
\begin{equation}
  {\mu_{\tr X^2}}^{\hat{\mathcal{O}}_{[2,(2,0)],r,\beta}} = c_r {\mu_{\tr Z^2}}^{\hat{\mathcal{O}}_{[2,(2,0)],0,\beta}}\eqndot
  \label{eq:mu X2 Z2 relation}
\end{equation}
An explicit calculation shows that $c_0 = -1/2$.

\bibliographystyle{utphys2}
\bibliography{TwoPointPaper}

\end{document}